 \documentclass[conference,compsoc]{IEEEtran}
\usepackage{graphicx}
\usepackage[noend]{algpseudocode}

\makeatletter
\def\theHALG@line{\thealgorithm.\arabic{ALG@line}}
\makeatother

\makeatletter
\newenvironment{suppfigure}{\par\begingroup\def\@captype{figure}}{\par\endgroup}
\newenvironment{supptable}{\par\begingroup\def\@captype{table}}{\par\endgroup}
\makeatother
\usepackage{amsmath,amssymb}
\usepackage[hidelinks]{hyperref}
\usepackage{booktabs}
\usepackage{comment}
\usepackage{cite}
\usepackage{xcolor} 
\newif\iforimarks
\orimarksfalse  % all \ori{} blocks have been integrated into the running text

\newif\ifrevmarks
\revmarksfalse

\usepackage{array}
\newcommand{\Description}[1]{}
\newcounter{algorithm}
\newenvironment{algorithm}[1][]{%
  \refstepcounter{algorithm}\begin{figure}[#1]\small%
}{\end{figure}}
\newcommand{\algcaption}[1]{\noindent\textbf{Algorithm \thealgorithm: #1}\par\smallskip}

\newcommand{\safeincludegraphics}[2][]{%
  \IfFileExists{#2}{%
    \includegraphics[#1]{#2}%
  }{%
    \fbox{\scriptsize Figure file unavailable}%
  }%
}

\title{Binarization as an Attack Surface in Weightless Neural Networks}
\author{
\IEEEauthorblockN{
Lucas Senos Coutinho\IEEEauthorrefmark{1},
Matheus Vargas\IEEEauthorrefmark{1},
Pedro Teixeira\IEEEauthorrefmark{1}\IEEEauthorrefmark{5},
Heudson  Tosta Mirandola\IEEEauthorrefmark{1},\\
Ori Rottenstreich\IEEEauthorrefmark{2},
Priscila Machado Vieira  Lima\IEEEauthorrefmark{1},
Leandro Santiago\IEEEauthorrefmark{3},
Daniel  Sadoc Menasche\IEEEauthorrefmark{1}
}
\IEEEauthorblockA{\IEEEauthorrefmark{1} Universidade Federal do Rio de Janeiro (UFRJ), Rio de Janeiro, Brazil }
\IEEEauthorblockA{\IEEEauthorrefmark{2} Technion--Israel Institute of Technology, Haifa, Israel\\ }
\IEEEauthorblockA{\IEEEauthorrefmark{3} Universidade Federal Fluminense (UFF), Niterói, Brazil  \\
\IEEEauthorrefmark{5}  Elytron Cybersecurity  }
}
\date{}

\begin{document}

\maketitle

\begin{comment}
\begin{abstract}
Weightless Neural Networks (WNNs) are lightweight learning models based on binarization, hashing, and memory lookup, making them attractive for resource-constrained classification and intrusion detection. Their discrete computation, however, does not by itself imply adversarial robustness. We study WiSARD-family models by isolating the mechanism through which a continuous perturbation reaches the classifier: it must cross an encoding threshold, change at least one binary tuple, and redirect a memory query. We formalize this mechanism with a threshold-crossing bound and a response-margin certificate for standard Boolean WiSARD. Controlled MNIST experiments then compare linear and data-driven thermometer encodings under model-agnostic sensitivity probes, a score-based Simulated Annealing attack, and surrogate-transfer attacks. The matched-architecture comparisons show that encoding granularity can trade a small clean-accuracy loss for a large decrease in attack success. Experiments on Bot-IoT, CICIDS, and UNSW-NB15 further show that aggregate accuracy can conceal missed attacks and false alarms.
%Overall, we bring awareness about   binarization is part of the attack surface  and should be selected jointly with model capacity and operational risk.
\end{abstract}
\end{comment}

\begin{abstract}
Weightless Neural Networks (WNNs) are lightweight classifiers that binarize their inputs and learn by storing binary patterns in lookup memories, which makes them attractive for resource-constrained tasks such as intrusion detection. Because a small perturbation that crosses no binarization threshold leaves the binary input unchanged, WNNs may appear naturally robust to adversarial examples; a perturbation that does cross a threshold, however, can still change the prediction. We study this tension in WiSARD, the canonical RAM-based WNN, and in two variants that replace its memories with Bloom filters, Bloom WiSARD and BTHOWeN. A continuous perturbation reaches these classifiers only by crossing an encoding threshold, changing at least one binary tuple, and redirecting a memory query; we formalize this mechanism with a threshold-crossing bound and a response-margin certificate for standard Boolean WiSARD. Controlled MNIST experiments compare linear and data-driven thermometer encodings under model-agnostic sensitivity probes, a score-based Simulated Annealing attack, and surrogate-transfer attacks. Changing only the encoding while keeping each model otherwise fixed shows that encoding granularity can trade a small clean-accuracy loss for a large decrease in attack success. On the Bot-IoT, CICIDS, and UNSW-NB15 intrusion-detection datasets, adversarial perturbations lead WNN detectors to classify many attacks as normal traffic and to raise false alarms, even when their aggregate accuracy remains high.
\end{abstract}

\section{Introduction}

Weightless Neural Networks (WNNs) are lightweight learning models that replace
floating-point multiply--accumulate operations with binarization, hashing, and
memory lookups~\cite{bacellar2024differentiable,carneiro2019exact,WiSARD}. This makes them attractive for resource-constrained networking
applications such as traffic classification, anomaly detection, and intrusion
detection~\cite{8702797,BTHOWeN}. Their design is also closely related to Bloom
filters, which are widely used in networked systems for compact approximate
membership tests in settings such as routing, caching, distributed systems,
intrusion detection, and blockchain networks~\cite{barbosa2025updating,bloom_filter,
reviriego2021analyzing,10458880,
banoun2026learning,
almashaqbeh2024adversary,
lotan2025adversarially,
gerbet2015power}. In both cases, efficiency
comes from compact binary representations and hash-based access to memory.

Despite these advantages, the adversarial robustness of WNNs remains poorly
understood. This is a critical gap: machine-learning systems deployed in
security-sensitive networking environments may be exposed to inputs crafted to
induce misclassification. In conventional deep neural networks, such
adversarial examples have been extensively studied~\cite{pelekis2025adversarial,intriguing_properties};
for WNNs, however, the attack surface is different. Perturbations do not
directly exploit differentiable weights. Instead, they may change the binary
encoding of the input, redirect tuple addresses, alter RAM or Bloom-filter
responses, and ultimately change the discriminator selected by the model.

We study this gap through the input binarizer, the first non-smooth component
of a WiSARD-family pipeline. A perturbation that crosses no encoding threshold
leaves the binary input unchanged and is   invisible to the classifier.
Once thresholds are crossed, however, tuple addresses and RAM or Bloom-filter
queries may change. Discreteness therefore creates locally invariant regions,
but it does not provide a global robustness guarantee. The size and placement
of those regions depend on the encoding, while the effect of leaving a region
also depends on tuple construction, memory occupancy, hash collisions, and
bleaching~\cite{bbleaching}.  We refer to this
chain of 1) threshold crossing, 2) tuple change, and 3) memory-query redirection as
the \emph{threshold-crossing mechanism}.

We evaluate this mechanism in two settings: MNIST and Intrusion-Detection Systems (IDS). On MNIST~\cite{deng2012mnist}, model-agnostic random
perturbations first measure encoding sensitivity; such perturbations are explicitly treated
as probes rather than adversarial attacks. A score-based Simulated Annealing
attack and attacks transferred from a differentiable surrogate then test
whether the locally invariant regions can be crossed deliberately. 

We study DictWiSARD, our implementation of standard Boolean
WiSARD~\cite{WiSARD} (hereafter Standard WiSARD), whose memory nodes are exact,
collision-free binary RAMs. Varying the encoding granularity of DictWiSARD
while keeping its architecture fixed changes only the binarizer, and
therefore provides the strongest evidence that the encoding itself drives
robustness.

A second controlled comparison changes   the input encoding of
BTHOWeN~\cite{BTHOWeN}, a Bloom-filter-based WNN, from its original Gaussian
thermometer to a linear one, keeping all other parameters fixed. We also
compare DictWiSARD, BTHOWeN, and Bloom WiSARD~\cite{bloom_WiSARD}, noting that   these
models differ in capacity, hashing, and bleaching~\cite{bbleaching}.   Finally, we evaluate Standard WiSARD and Bloom
WiSARD on three IDS datasets---Bot-IoT~\cite{bot_iot}, CICIDS~\cite{cicids2017},
and UNSW-NB15~\cite{unsw_nb15}---under binary and multiclass classification. % These normalized-feature experiments are useful for classifier stress testing.

This paper makes the following contributions.

\textbf{A threshold-crossing account of WNN robustness.}
We formalize how a norm-bounded perturbation changes thermometer bits and tuple
addresses, and derive a response-margin robustness certificate for Standard WiSARD. This
separates properties implied by the encoding and classifier structure from
empirical observations about particular attacks (Section~\ref{sec:methodology}).

\textbf{Controlled and adaptive empirical evidence.}
We distinguish model-agnostic sensitivity probes from score-based and
surrogate-transfer attacks. Within-model experiments show that increasing
thermometer resolution can substantially raise attack success even when clean
accuracy changes little (Sections~\ref{sec:experimental-methodology} and~\ref{sec:experimental-results}). %These results demonstrate vulnerability, not certified robustness of the coarser configurations.

\textbf{IDS stress tests.}
Across Bot-IoT, CICIDS, and UNSW-NB15, we report Matthews Correlation Coefficient (MCC), malicious-to-Normal
evasion rate,  False Alarm Rate (FAR), and class-level miss rates. This exposes failure
modes hidden by accuracy while making the limitations of normalized-feature
perturbations explicit (Section~\ref{sec:ids_results}).

The remainder of the paper is organized as follows.
Section~\ref{sec:related-background} reviews WNNs, Bloom-filter-based WNNs, and
adversarial-learning background. Section~\ref{sec:methodology} introduces the
binarization schemes and robustness analysis. Section~\ref{sec:experimental-methodology}
details the attack methodology. 
Sections~\ref{sec:experimental-results} and~\ref{sec:ids_results} present  the MNIST and IDS experimental results,
Section~\ref{sec:engineering_implications} discusses their implications, and
Section~\ref{chap:concl} concludes the paper.

\section{Related Work and Background}
\label{sec:related-background}

The adversarial robustness of WNNs~\cite{WiSARD,jadhao2025hybrid,bacellar2026arithmetic} remains
largely unexplored, but closely related ideas appear in the
adversarial-learning literature on input quantization.
% Space recovery: the opening two sentences repeated the first paragraph of
% the Introduction almost verbatim.
For example, thermometer encoding has been proposed as a
defense mechanism because discretization can disrupt gradients used by
attackers~\cite{thermometer,buckman2018thermometer}. Madry et al. also observed
that adversarially trained networks may learn threshold-like behavior in early
layers~\cite{madry2019deeplearningmodelsresistant}. These observations motivate
our study: since WNNs rely on explicit binarization as part of their native
computation, input quantization may provide stability against some
perturbations, but it may also introduce threshold-crossing vulnerabilities.

 WNNs~\cite{ludermir1999weightless} were
popularized by WiSARD~\cite{WiSARD}. In contrast to Deep Neural Networks (DNNs), WiSARD does not learn
real-valued weights. Instead, each class is represented by a discriminator
composed of RAM-based neurons. Given a binarized input with $I$ bits and RAM
nodes addressed by tuples of size $n$, each discriminator contains
$N$ RAM nodes. In the simplest
non-overlapping construction, $N=I/n$, 
and a dense implementation requires  $|\mathcal{C}|\,N2^n$ memory entries,
where $\mathcal{C}$ is the set of classes.  The implementations evaluated here
allow the tuple mapping and number of memory nodes to be configured
independently, as specified in Table~\ref{tab:params}.

During training, each binarized input is mapped to
tuples through a fixed pseudo-random mapping and modifies only   the
discriminator associated with its true class. During inference, the same input
is queried against all discriminators, and the predicted label is the class
whose discriminator obtains the largest response.  Thus, a perturbation can affect the WiSARD response only if it changes the binary encoding and thereby redirects at least one tuple to a different memory address; however, such redirection does not necessarily change the final prediction.

\subsection{Bloom-filter-based WNNs}
\label{subsec:bloom-based-wnns}

The main scalability limitation of WiSARD is the exponential memory growth in
the tuple size $n$. For example, $n=64$ would require $2^{64}$ possible
addresses for a single RAM node. Bloom WiSARD addresses this limitation by
replacing RAM nodes with Bloom-filter-based memories~\cite{bloom_WiSARD}. A
Bloom filter is a compact hash-based data structure for approximate membership
queries~\cite{bloom_filter}: under the standard model, it has no false
negatives, but may return false positives depending on the filter size, the
number of hash functions, and the number of inserted elements.

Before presenting the algorithms, we introduce the basic notation, summarized
in Table~\ref{tab:notation}.  The binarized input is denoted by $z$. The WiSARD mapping divides $z$ into
tuples, where each tuple $z_t$ contains a subset of the input bits   and is indexed by $t \in \mathcal{T}$. For each class discriminator, every
tuple index $t$ has an associated storage unit, or memory node, that records 
patterns observed for tuple $z_t$.  In Standard WiSARD, this memory node is an
exact RAM table: the tuple bits determine an address, and training sets the
corresponding entry to 1. In Bloom WiSARD, the memory node is implemented as a
Bloom filter: the tuple bits are mapped by multiple hash functions, and training
sets the corresponding Bloom-filter positions to~1.

We denote by
$\mathcal{M}_{c,t}$ the memory node associated with class $c$ and tuple $t$.
For each tuple $t$, the set $\mathcal{F}_t$ contains the addressing or hash
functions used to access $\mathcal{M}_{c,t}$. During inference, $R_c$ is the
total response of class discriminator $c$, and $h_t$ is the temporary response
of tuple $t$, i.e., the number of active memory positions reached by the
functions in $\mathcal{F}_t$.

Algorithms~\ref{alg:unified-train} and~\ref{alg:unified-test} summarize this
unified view of WiSARD and Bloom WiSARD. During training, the input is processed
one tuple at a time (line~\ref{line:train-tuple-loop}). Each tuple is
mapped through its associated addressing or hash functions (line~\ref{line:train-function-loop}). The selected memory positions are then set
to 1 in the memory node of the true class (line~\ref{line:train-store}). Thus, training stores class-specific binary
patterns: exact tuple addresses in standard WiSARD and Bloom-filter hash
positions in Bloom WiSARD.

\begin{algorithm}[t]
\algcaption{Training for WiSARD/Bloom WiSARD}
\label{alg:unified-train}
\begin{algorithmic}[1]
\Require Binarized training example $z$, label $y$, tuple set $\mathcal{T}$,
  per-tuple function sets $\mathcal{F}_t$, and class memories $\mathcal{M}_{c,t}$.
\Ensure Trained model, i.e., updated memories. 
\For{each tuple $t\in\mathcal{T}$}\label{line:train-tuple-loop}
  \For{each function $f\in\mathcal{F}_t$}\label{line:train-function-loop}
    \State $\mathcal{M}_{y,t}[f(z_t)]\gets 1$\label{line:train-store}
  \EndFor
\EndFor
\end{algorithmic}
\end{algorithm}

\begin{algorithm}[t]
\algcaption{Inference for WiSARD/Bloom WiSARD}
\label{alg:unified-test}
\begin{algorithmic}[1]
\Require Binarized input $z$, classes $\mathcal{C}$, tuple set $\mathcal{T}$,
  per-tuple function sets $\mathcal{F}_t$, and memories $\mathcal{M}_{c,t}$.
\Ensure Predicted class.
\For{each class $c\in\mathcal{C}$}\label{line:infer-class-loop}
  \State $R_c\gets 0$\label{line:infer-init-class}
  \For{each tuple $t\in\mathcal{T}$}\label{line:infer-tuple-loop}
    \State $h_t\gets 0$\label{line:infer-init-hit}
    \For{each function $f\in\mathcal{F}_t$}\label{line:infer-function-loop}
      \State $h_t\gets h_t+\mathcal{M}_{c,t}[f(z_t)]$\label{line:infer-accumulate}
    \EndFor
    \If{$h_t=|\mathcal{F}_t|$}\label{line:infer-membership}
      \State $R_c\gets R_c+1$\label{line:infer-increment}
    \EndIf
  \EndFor
\EndFor
\State \Return $\arg\max_{c\in\mathcal{C}} R_c$\label{line:infer-return}
\end{algorithmic}
\end{algorithm}

During inference, the model computes one response $R_c$ for each class
(line~\ref{line:infer-class-loop}). For each class, the response is initialized
to zero (line~\ref{line:infer-init-class}) and then accumulated over the input
tuples (line~\ref{line:infer-tuple-loop}). For a given tuple, the algorithm
counts how many associated memory positions are active
(line~\ref{line:infer-accumulate}). The tuple
contributes to the class response only if all associated positions are active
(line~\ref{line:infer-membership}). Finally, the
predicted class is the one with the largest discriminator response
(line~\ref{line:infer-return}).

 Standard WiSARD, a classical Bloom filter, and Bloom WiSARD are recovered as
special cases of this framework through the choice of $\mathcal{C}$,
$\mathcal{T}$, $\mathcal{F}_t$, and the memory size:
\begin{itemize}
  \item \textbf{Standard WiSARD:} $\mathcal{F}_t=\{f_t\}$ contains a single
  injective function that reads the $n$ tuple bits as a binary address, and
  $\mathcal{M}_{c,t}$ has $2^n$ entries. Each tuple pattern therefore has its
  own entry, and the test $h_t=|\mathcal{F}_t|=1$
  (line~\ref{line:infer-membership}) reduces to reading that entry.
  \item \textbf{Classical Bloom filter:} with a single class, a single tuple
  containing all input bits ($|\mathcal{C}|=|\mathcal{T}|=1$), and
  $\mathcal{F}_t=\{f_{t,1},\dots,f_{t,r}\}$ a set of $r$ hash functions into an
  $m$-bit memory, line~\ref{line:train-store} inserts $z$ into the filter, and
  the test $h_t=r$ (line~\ref{line:infer-membership}) is the standard
  membership query: it never misses an inserted input, but may return false
  positives when other inputs have set the same $r$ positions ($r$ is \texttt{n\_hash} in Table~\ref{tab:params}).
  \item \textbf{Bloom WiSARD:} each tuple of each class uses its own such
  filter, with $m\ll 2^n$, so the class response $R_c$ counts the tuples whose
  patterns the filter reports as seen. With $r=1$, $m=2^n$, and the identity
  address map, Bloom WiSARD reduces to Standard WiSARD.
\end{itemize}

This formulation highlights where adversarial perturbations can affect the
model. A perturbation that changes the binarized input may change a tuple
$z_t$, which in turn changes the addressed memory locations
(lines~\ref{line:train-store} and~\ref{line:infer-accumulate}). In Bloom
WiSARD, such changes interact with hash collisions and Bloom-filter false
positives: a perturbed tuple may fail to activate positions associated with the
true class, or it may activate positions in a competing class. Thus, collisions
are not only an efficiency issue; they also affect discriminator responses and
therefore form part of the model's adversarial behavior.

BTHOWeN (Bleached, Thermometer-Encoded, Hashed-input Optimized Weightless
Neural Network) further extends this Bloom-filter-based design
space~\cite{BTHOWeN}. It combines counting Bloom filters, bleaching, hashing,
and Gaussian thermometer encoding to improve accuracy and hardware efficiency.
With bleaching~\cite{bbleaching}, repeated tuple insertions increment counters,
and a tuple is considered active at inference time only if all its mapped
positions reach a threshold $\beta$. This suppresses rare patterns and can improve
clean-data behavior, but it also changes the response surface seen by an
attacker. Consequently, binarization thresholds, hash collisions, Bloom-filter
false positives, and bleaching thresholds all contribute to the robustness
trade-offs analyzed in this work.

\begin{table}[t]
\caption{Notation.}
\label{tab:notation}
\small \resizebox{\columnwidth}{!}{
\begin{tabular}{@{}lp{0.74\columnwidth}@{}}
\toprule
\textbf{Symbol} & \textbf{Meaning} \\
\midrule
$x=(x_1,\ldots,x_d)$ & Input \emph{before} binarization (image or feature vector), with $d$ features indexed by $j$ \\
$b$ & \texttt{bits\_per\_input}: thermometer bits per feature \\
$\Theta_j=(\tau_{j,1},\ldots,\tau_{j,b_j})$ & Thermometer thresholds of feature $j$ \\
$z=B(x)$ & Input \emph{after} binarization: the concatenated thermometer code \\
$t\in\mathcal{T}$, $z_t$ & Tuple index and the bits of $z$ in tuple $t$ \\
$n$ & \texttt{tuple\_size}: bits of $z$ read by each memory node \\  
$c\in\mathcal{C}$, $y$ & Class index and true class \\
$\mathcal{M}_{c,t}$ & Memory node of class $c$ for tuple $t$ \\
$\mathcal{F}_t$, $f$ & Addressing or hash functions of tuple $t$; $r$ hashes into $m$ positions for Bloom WiSARD \\
$h_t$, $R_c$ & Active positions reached by tuple $t$; response of class $c$ \\
% $g$ & Classifier (full pipeline, from $x$ to a predicted class) \\
$\delta$, $\epsilon$ & Perturbation and its budget, $\|\delta\|_p\leq\epsilon$ \\ 
\bottomrule
\end{tabular} }
\end{table}

\subsection{Adversarial Machine Learning Background}
\label{subsec:aml-background}

Adversarial attacks aim to induce misclassification by applying a small
perturbation to an input while preserving its semantic class. Let $g$ be a
classifier, $x$ an input correctly classified as $y$, and $\delta$ a
perturbation. An adversarial example is an input
$x_{\mathrm{adv}} = x+\delta$ such that $g(x_{\mathrm{adv}})\neq y$,
subject to a perturbation budget $\|\delta\|_p \leq \epsilon$. Common choices
include the $L_2$ norm,
$\|\delta\|_2=\sqrt{\sum_{j=1}^{d}\delta_j^2}$, which bounds the Euclidean
magnitude of the perturbation, and the $L_\infty$ norm,
$\|\delta\|_\infty=\max_{1\leq j\leq d}|\delta_j|$, which bounds the maximum change applied
to any individual feature or pixel~\cite{survey_attacks_defenses}. In image
classification, this constraint is often interpreted as preserving visual
similarity; in networking and tabular settings, it must also respect the
feature domain and preprocessing pipeline.

This work focuses on evasion attacks, where the model is already trained and
the adversary modifies inputs presented during inference to cause incorrect predictions. We do
not consider poisoning or backdoor attacks, where the training data or training
procedure is compromised. The threat model is shaped by the non-differentiable
nature of WNNs: binarization, memory lookup, hashing, and discriminator
aggregation do not provide gradients in the same way as conventional neural
networks. Therefore, direct white-box gradient attacks are not   
applicable to the target WNN.

We consider three complementary attack families. Random perturbations provide a
baseline for measuring sensitivity to norm-constrained noise. Query-based
optimization attacks treat the WNN as a black-box scoring function and search
for perturbations that reduce the true-class response. Transfer-based attacks
train a differentiable surrogate model to mimic the WNN, generate adversarial
examples using gradient-based attacks on the surrogate, and then evaluate
whether these examples transfer to the original WNN
\cite{practical_black_box,papernot2016transferabilitymachinelearningphenomena}.
This setting is especially relevant because it tests whether a
non-differentiable, memory-based classifier can still be attacked through a
smooth approximation of its input-output behavior.

% For WNNs, adversarial robustness depends not only on the classifier but also on the binarization function. A perturbation that does not cross any binarization threshold leaves the binary representation unchanged and is invisible to the model. Conversely, a perturbation that crosses one or more thresholds can change tuple addresses, alter RAM or Bloom-filter responses, and modify the final discriminator scores. This makes input quantization a central part of the attack surface analyzed in this work.

\section{Binarization and Robustness}
\label{sec:methodology}

In this section we describe the
binarization schemes, since they determine how continuous perturbations are
translated into binary tuple-address changes.  Then, we introduce a simple threshold-crossing bound to assess the robustness of WNNs with respect to input encodings. 

\subsection{Input Binarization}
\label{sec:binarization}

Input binarization is the first step in WNN inference and a central design
choice for both clean accuracy and robustness. Since WNNs store and query binary
patterns, a perturbation in the continuous input affects the model only if it
crosses one or more binarization thresholds and changes the resulting binary
code. The models evaluated in this work use thermometer-style encodings:
WiSARD variants use linear thermometer encoding, whereas BTHOWeN uses a
data-driven Gaussian thermometer encoding.

\textbf{Linear thermometer encoding. }
Linear thermometer encoding discretizes the normalized range $[0,1]$ using
evenly spaced thresholds~\cite{thermometer}. Given
\texttt{bits\_per\_input} $b$, the encoding creates $b$ thresholds and compares
each input value against them. For example, with $b=4$, the thresholds are
the ordered sequence $\Theta=(0.2,0.4,0.6,0.8)$, so a value of $0.7$ is encoded as $1110$.  Each bit tests non-strict inequality,
$\mathbf{1}\{x_j\geq\tau_{j,k}\}$, and equality is therefore encoded as $1$.

This encoding is
relatively stable when $b$ is small, because small perturbations often do not
cross any threshold, and even when they do, they usually change only a few bits.

\textbf{Gaussian thermometer encoding. }
BTHOWeN uses a data-driven Gaussian thermometer encoding~\cite{BTHOWeN}. Instead
of spacing thresholds uniformly, it estimates the mean and standard deviation of
each feature from the training data and places thresholds according to Gaussian
quantiles. This concentrates resolution near high-density regions of the data
distribution, which can improve clean accuracy. However, it can also increase
adversarial sensitivity: when several thresholds are close together, a small
perturbation near the mean may cross multiple thresholds and produce a large
change in the binary representation.

\subsection{Threshold-crossing Bound}
\label{sec:threshold_crossing_bound}

This section quantifies how far a bounded perturbation can propagate through a
WNN: from thermometer thresholds, to binary code bits, to memory tuples, and
finally to class responses. The result is a sufficient condition under which
no perturbation within the budget can change the prediction. 

Let $x=(x_1,\ldots,x_d)$, where $j\in\{1,\ldots,d\}$ indexes an input
feature and $x_j$ is its value. For each feature $j$, let
$\Theta_j=(\tau_{j,1},\ldots,\tau_{j,b_j})$ be its ordered threshold sequence,
where $k\in\{1,\ldots,b_j\}$ indexes a threshold within that feature, and let
$B(x)$ denote the concatenated thermometer code, $z=B(x)$. 
For a perturbation
$\delta$ satisfying $\|\delta\|_\infty\leq\epsilon$, define
 $H_\epsilon(x)$ as the \emph{threshold exposure} of $x$ at budget
$\epsilon$: the number of thresholds that lie within $\epsilon$ of the input,
\begin{equation}
H_\epsilon(x)=
\sum_j\sum_{k=1}^{b_j}
\mathbf{1}\{|x_j-\tau_{j,k}|\leq\epsilon\}.
\label{eq:threshold-crossing-bound}
\end{equation}
For example, let $x=(0.34,0.74)$,
$\Theta_1=\Theta_2=(0.2,0.4,0.6,0.8)$, and $\epsilon=0.07$. The interval
$[0.27,0.41]$ for feature 1 contains only the threshold $0.4$, while the
interval $[0.67,0.81]$ for feature 2 contains only $0.8$. Therefore,
$H_{0.07}(x)=1+1=2$, so at most two thermometer bits can change.

For coordinate $j$, the $L_\infty$ constraint keeps $x_j+\delta_j$ inside
$[x_j-\epsilon,x_j+\epsilon]$. A threshold outside this interval remains on
the same side of the feature value before and after perturbation, so its
comparison bit cannot change; counting the thresholds inside all such
intervals therefore upper-bounds the number of changed bits. Hence, 
\begin{equation}
d_H\!\left(B(x),B(x+\delta)\right)\leq H_\epsilon(x),
\label{eq:hamming-bound}
\end{equation}
where $d_H$ is Hamming distance. In particular, $H_\epsilon(x)=0$ is an
input-level certificate: every perturbation in the $L_\infty$ ball produces
exactly the same WNN input code. Because Equation~\eqref{eq:threshold-crossing-bound}
uses closed threshold windows, the containment in Equation~\eqref{eq:hamming-bound}
is unaffected by the tie convention used when an input is exactly equal to a
threshold.

The WiSARD mapping assigns encoded bits to memory tuples. Let
$Q_\epsilon(x)$ be the set of tuples containing at least one bit counted by
$H_\epsilon(x)$; then at most $q=|Q_\epsilon(x)|$ tuple addresses can change.
Equations~\eqref{eq:threshold-crossing-bound}--\eqref{eq:hamming-bound} depend
only on the encoding and therefore apply to every model in this paper; the next
step depends on the memory. Let $y$ be the clean predicted class (the true
class, for a correctly classified input), and define its \emph{response margin}
as
\begin{equation}
\Delta(x)=R_y(B(x))-\max_{c\neq y}R_c(B(x)). \label{eq:deltax}
\end{equation}
In Standard WiSARD, each changed tuple can alter a class response by at most
one, so under any $\|\delta\|_\infty\leq\epsilon$ the response of $y$ decreases
by at most $q$ while a competitor increases by at most $q$. This yields the
\emph{response-margin certificate}:\begin{equation}
\Delta(x)>2q \;\Longrightarrow\;
\arg\max_{c}R_c\bigl(B(x+\delta)\bigr)=y, \; \forall\,\|\delta\|_\infty\leq\epsilon.
\label{eq:response-margin}
\end{equation}
In words, if the clean class leads every competitor by more than twice the
number of tuples that can change, then no perturbation within the budget can
change the prediction: even if every changeable tuple both lowers the
response of $y$ and raises that of its strongest competitor, $y$ still wins.

The    response-margin certificate  
 in Equation~\eqref{eq:response-margin}  is established here for  
Standard WiSARD~\cite{WiSARD}. Extensions accounting for Bloom collisions in Bloom WiSARD~\cite{bloom_WiSARD}, counting memories~\cite{counting_bloom},
and bleaching~\cite{bbleaching} remain future work.

The rest of the paper uses the above bound   in two ways. Threshold exposure
explains why linear and Gaussian encodings react so differently to random
probes (Section~\ref{sec:binarization_impact} and
Figure~\ref{fig:bin_distance}), and the bound in
Equation~\eqref{eq:hamming-bound} predicts that attack success grows with
\texttt{bits\_per\_input} (Sections~\ref{subsec:sa_results}
and~\ref{sec:surrtransfres}). %; and the response-margin certificate motivates the threshold-placement defenses outlined as future work in  Section~\ref{chap:concl}.
% [CR01] ORIGINAL - Alcance do certificado
% This certificate is local and conservative. Counting
% memories, bleaching, and Bloom collisions require occupancy-aware bounds; the
% empirical results below study those effects.
% [CR01] REVISED (Ori) -- integrated into the paragraph above; future-work
% sentence moved to Section~\ref{chap:concl}.
% \ori{The above bit-level bound depends
% only on encoding; the response-margin guarantee is established here for exact
% Boolean WiSARD. Extensions accounting for Bloom collisions, counting memories,
% and bleaching remain future work.}
% [CR01] END

% Appendix~\ref{app:threshold_certificate} provides a direct MNIST diagnostic of Equations~\eqref{eq:threshold-crossing-bound}--\eqref{eq:response-margin} for exact Boolean DictWiSARD, measuring the bit-level bound, the number of potentially affected tuples, and the fraction of clean-correct samples that satisfy the response-margin certificate.

\begin{figure}[t]
    \centering
\begin{tabular}{ccc}\safeincludegraphics[width=0.25\linewidth]{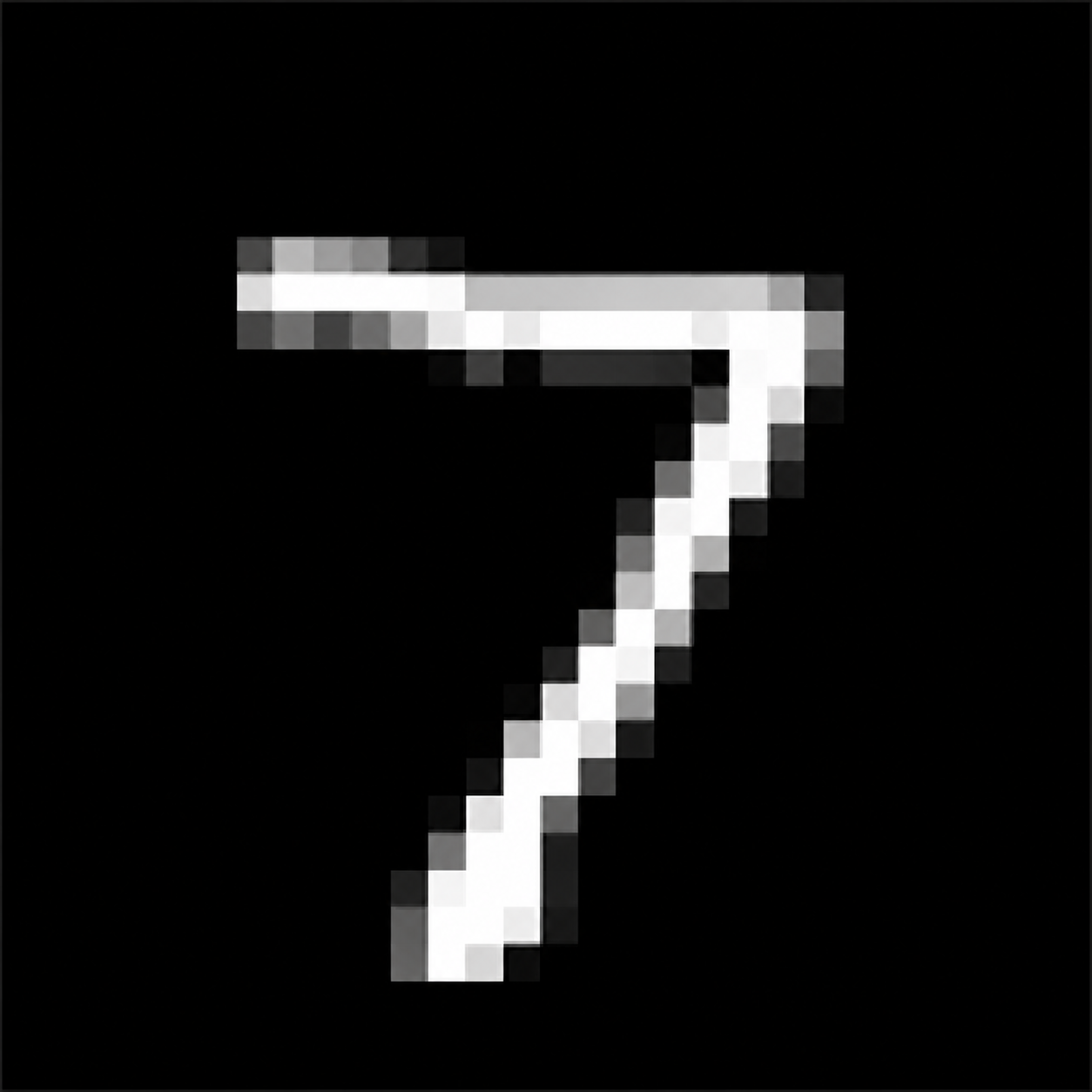} &  
\safeincludegraphics[width=0.25\linewidth]{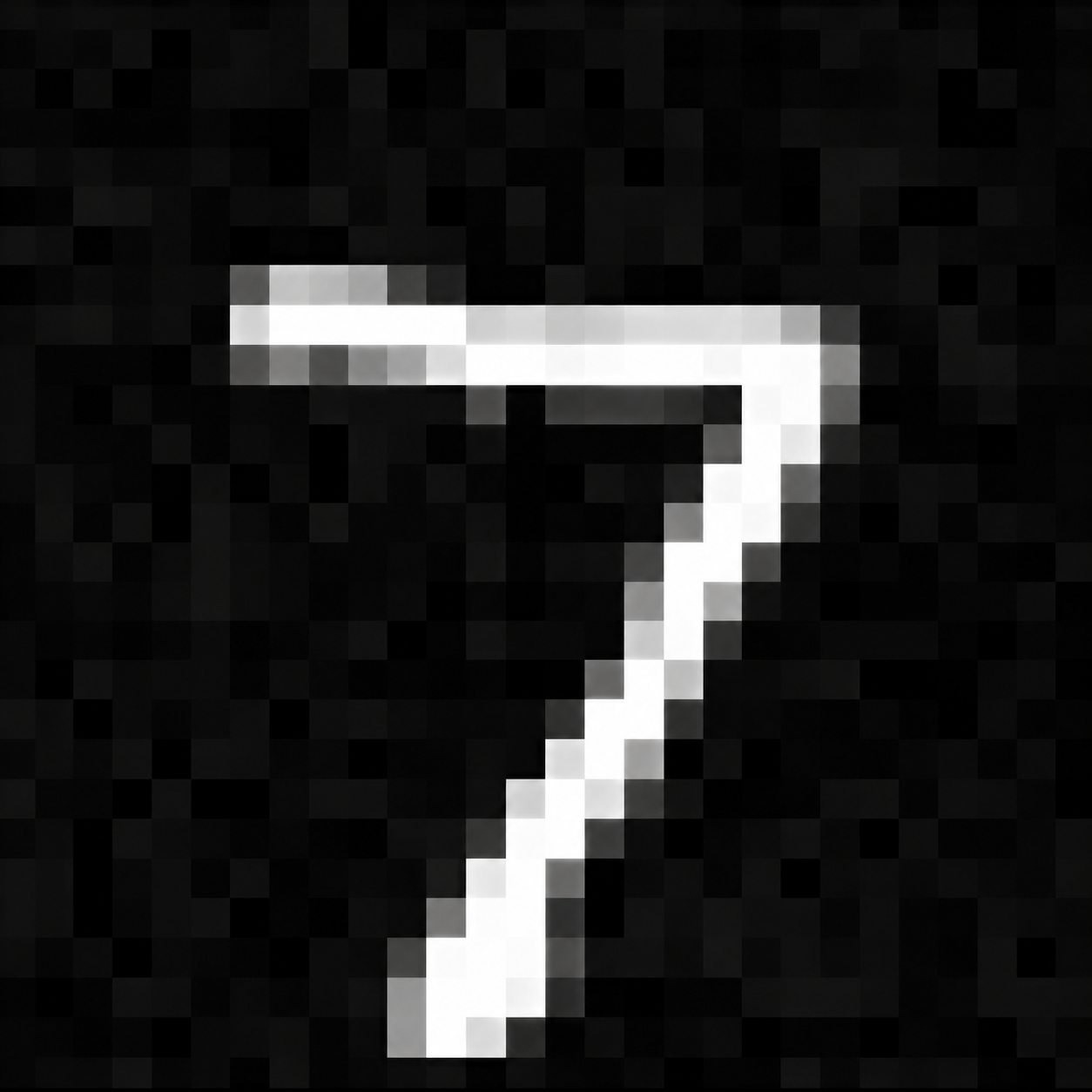} & 
\safeincludegraphics[width=0.25\linewidth]{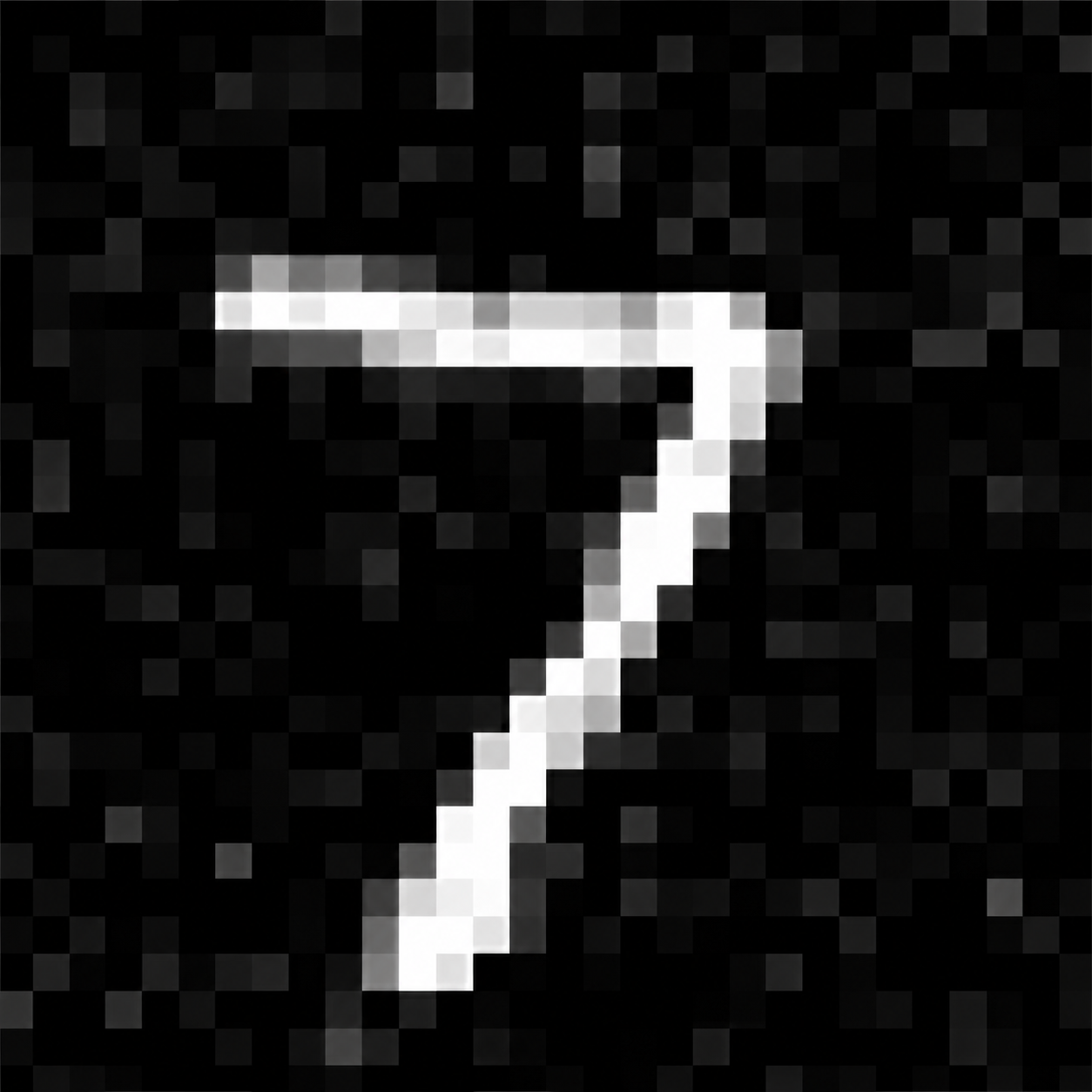} \\ 
(a)  & (b)  & (c)
\end{tabular} 
    \caption{Example of random perturbations applied to an MNIST digit: (a) Original; (b) Random $L_{\infty}$; (c) Random $L_2$.}
\Description{Visualization for the random attacks on an instance of the number 7 of the MNIST.}
    \label{fig:random_attacks}
\end{figure}

\section{Attack Methodology}
\label{sec:experimental-methodology}

This section describes the experimental methodology used to evaluate the
adversarial robustness of WiSARD-family WNNs.  In the following three subsections, we   present the random,
query-based, and surrogate-transfer attacks. %, followed by the evaluated models and the metrics used in the intrusion-detection experiments.

\subsection{Model-Agnostic Sensitivity Probes}
\label{subsec:random_attack}

We use two random perturbation attacks as baseline sensitivity tests. The
Random $L_\infty$ attack independently adds or subtracts $\epsilon=0.1$ to each
pixel with uniform probability and clips the result to $[0,1]$. The Random
$L_2$ attack samples Gaussian noise, normalizes it to have $L_2$ norm
$\epsilon=3$, adds it to the image, and clips the result to $[0,1]$. These
budgets follow common MNIST-scale adversarial settings~\cite{survey_attacks_defenses}.
The goal of these attacks is not to optimize against the model, but to measure
how often simple norm-constrained noise crosses binarization thresholds and
alters discriminator responses.  Figure~\ref{fig:random_attacks} illustrates their visual effect.

\subsection{Simulated Annealing Attack}
\label{subsec:sa_attack}

We use Simulated Annealing (SA) as a query-based black-box attack against
DictWiSARD. The attack searches for an $L_2$-bounded perturbation that reduces
the true-class discriminator response relative to the strongest competing
class. Let $x$ be a clean MNIST image and $x'$ a candidate
within the permitted $L_2$ ball. For an untargeted attack, the energy function
is the response margin of Equation~\eqref{eq:deltax} evaluated at the
candidate and normalized by the total response, with $y$ the true class,
\begin{equation}
  E(x') = \frac{\Delta(x')}{\sum_{c\in\mathcal{C}}R_c(B(x'))}.
    \label{eq:energia}
\end{equation}
Since the normalization is positive, the attack succeeds exactly when
$E(x')<0$, that is, when it drives the margin protected by the certificate of
Equation~\eqref{eq:response-margin} below zero.

The valid state space is
\[
    S =
    \{x' \in [0,1]^{784}
    \mid
    \|x' - x\|_2 \leq \epsilon\}.
\]
Transitions are designed to
target the WNN threshold-crossing mechanism: the attack selects a pixel at
random and moves it just across one of its two neighboring thermometer
thresholds, redrawing the move if it would exceed the $L_2$ budget. Budget spent
on a pixel is not recovered if the pixel later returns to its original bin,
which makes the attack conservative. This directly encourages threshold crossings and
therefore changes in the binary representation.

For the SA attack, we use the standard MNIST $L_2$ budget $\epsilon=3$. %The resulting rates are specific to this budget and should not be read as a budget-independent ranking. 
The SA procedure uses initial temperature $10$, exponential cooling by
a factor of $0.999$ per iteration down to a floor of $0.001$, and at most 5000
iterations per image, stopping early once the attack succeeds.
A proposed move from the current candidate $x'\in S$ to a new candidate
$x''\in S$ is accepted with probability
\[ P(x',x'') = \min\left(1,\exp\left(-\frac{E(x'')-E(x')}{T}\right)\right), \]
where $T$ is the current temperature. 

\subsection{Surrogate-Transfer Attack}
\label{subsec:surrogate_attack}

 The  surrogate-transfer attack consists in producing adversarial
examples   on a differentiable approximation of the WNN, and use them to  fool
the original non-differentiable model. % (see Section~\ref{subsec:surrogate_attack}). 
% We  evaluate transfer-based attacks using a differentiable surrogate model.
The target WNN is treated as a black box, and a Multi-Layer Perceptron (MLP) is
trained to mimic its output distribution. In the first stage, the MLP is trained
on the original MNIST inputs using soft labels obtained by querying the WNN,
with a Kullback--Leibler divergence objective between the surrogate and WNN
output distributions. In the second stage, gradient-based attacks are generated
against the surrogate and transferred back to the WNN.

We use FGSM, BIM, MI-FGSM, and PGD as implemented in
Foolbox~\cite{rauber2017foolbox}, with the corresponding standard
references~\cite{goodfellow2015explainingharnessingadversarialexamples, kurakin2017adversarialexamplesphysicalworld, 
dong2018boostingadversarialattacksmomentum,
madry2019deeplearningmodelsresistant}.

For the surrogate-transfer experiments, adversarial examples are constrained
by $\|\delta\|_\infty\leq 0.15$. We use the default Foolbox~3.3.4 attack
parameters. FGSM performs a single gradient-sign step. BIM and MI-FGSM use
10 iterations with step size $0.2\epsilon=0.03$ and no random initialization;
MI-FGSM additionally uses momentum factor $1.0$. PGD uses 40 iterations with
step size $\epsilon/30=0.005$ and random initialization within the
$L_\infty$ ball. All adversarial inputs are constrained to $[0,1]$.

The surrogate is an MLP with two hidden layers of 256 neurons each. We use this
simple architecture to approximate the score vector without reproducing the
target's discrete computation. All reported transfer rates are averaged over
three independent runs.

\section{Experimental Results with MNIST}
\label{sec:experimental-results}

In this section we consider experimental results with MNIST. The experiments proceed
in three steps. First, model-agnostic random perturbations characterize the
sensitivity of the   architectures and   binarization schemes to bounded
input changes (Section~\ref{sec:randomeval}). Second, controlled encoding comparisons isolate the
threshold-crossing mechanism itself: varying \texttt{bits\_per\_input} within
DictWiSARD, or replacing only the BTHOWeN binarizer, changes the encoding
while holding the rest of the model fixed, whereas comparisons across
DictWiSARD, Bloom WiSARD, and BTHOWeN also differ in hashing, memory
organization, capacity, and bleaching and are therefore descriptive (Section~\ref{sec:binarization_impact}). Finally,
query-based Simulated Annealing and surrogate-transfer attacks test whether an
adversary can deliberately exploit the threshold-crossing mechanism (Sections~\ref{subsec:sa_results} and~\ref{sec:surrtransfres}). Together,
these experiments distinguish stability to incidental perturbations from
robustness against adaptive attacks.
% [CR02]/[CR03] (Ori) integrated into this paragraph; the two future-work
% sentences ("more challenging image datasets", "future ablations") moved to
% Section~\ref{chap:concl}. Former subsections 5.1 and 5.2 merged below.

\subsection{\texorpdfstring{Models, Dataset, and Configurations}{Models, Dataset, and Configurations}}
\label{subsec:models_datasets}

The experiments use the 60,000 training and 10,000 test images of
MNIST~\cite{mnist}: $28\times28$ grayscale handwritten digits with pixel values
normalized to $[0,1]$. We evaluate the three architectures of
Section~\ref{subsec:bloom-based-wnns}. DictWiSARD implements Standard
WiSARD~\cite{WiSARD} with hash maps that store only the observed tuple
patterns, which is functionally equivalent to dense Boolean RAMs; Bloom
WiSARD~\cite{bloom_WiSARD,bloom_filter} and BTHOWeN~\cite{BTHOWeN} use Bloom-filter
memories. Table~\ref{tab:params} lists their parameters, using the notation of
Table~\ref{tab:notation} for \texttt{bpi} ($b$; also referred to as \texttt{bits\_per\_input}), and  \texttt{tuple\_size} ($n$). 
\texttt{RAMs} is the number of memory nodes per discriminator, \texttt{dtype}
the memory type, and \texttt{bleach} whether bleaching is used at inference. For Bloom-filter-based models, 
\texttt{filter\_size} ($m$), and \texttt{n\_hash} ($r$) denote the Bloom filter size and the number of hash functions, respectively.  

\begin{comment}
The experiments use 60,000 training and 10,000 testing grayscale images of
handwritten digits, each with size $28\times28$ and pixel values normalized to
$[0,1]$~\cite{mnist}. We evaluate three WNN architectures. DictWiSARD is a
dictionary-based implementation of Standard WiSARD, which stores only observed tuple patterns and is functionally equivalent to dense Boolean RAMs~\cite{WiSARD},
using hash maps to store observed tuple patterns. Bloom WiSARD replaces
dictionary memories with Bloom-filter-based memories, reducing memory usage at
the cost of controlled collisions~\cite{bloom_filter,bloom_WiSARD}. BTHOWeN is
a Bloom-filter-based WNN architecture that combines hashing, bleaching, counting
Bloom filters, and Gaussian thermometer encoding~\cite{BTHOWeN}.
Table~\ref{tab:params} summarizes the main model parameters:
\texttt{bpi} is the number of bits generated per input feature (\texttt{bits\_per\_input}), e.g.,  
number of bits per pixel for images;  \texttt{tuple\_size} denotes
the number of binarized input bits read by each memory node;  \texttt{RAMs} gives the number of memory nodes per discriminator,
\texttt{dtype} gives the memory type, and \texttt{bleach} indicates whether
bleaching is used at inference time. For Bloom-filter-based models,
\texttt{filter\_size} and \texttt{n\_hash} denote the Bloom filter size and the
number of hash functions, respectively.
\end{comment}
\setlength{\tabcolsep}{2.5pt}

\begin{table}[t]
\centering
\caption{Baseline WNN configurations.}
\Description{Weightless models' parameters for baseline performance evaluation.}   \footnotesize
\label{tab:params} 
\begin{tabular}{lccccccc}
\toprule
\textbf{Model} & \textbf{bpi} & \shortstack{\textbf{tuple\_}\\\textbf{size}} & \textbf{RAMs} & \textbf{dtype} & \textbf{bleach} & \shortstack{\textbf{filter\_}\\\textbf{size}} & \textbf{n\_hash} \\
\midrule
DictWiSARD & 4 & 48 & 50 & bool & false & - & - \\
Bloom WiSARD & 4 & 48 & 50 & bool & false & 37412 & 4 \\
BTHOWeN & 6 & 49 & 96 & int & true & 8192 & 4 \\
\bottomrule
\end{tabular}
\end{table}

% Include hyperparameters and execution environment of the experiments.

%To investigate the effect of binarization in isolation, we later will  evaluate variants of the DictWiSARD model with distinct \texttt{bits\_per\_input} parameter values.

% \end{comment}

%\section{Results and Analysis} \label{chap:results_and_analysis}

\subsection{Sensitivity to Model-Agnostic Perturbations} \label{sec:randomeval}

Random noise   provides a baseline for measuring how each WNN reacts
to simple norm-constrained perturbations. Although random noise does not optimize
against the model, it reveals a sharp contrast between the evaluated
binarization schemes.

Table~\ref{tab:random_attack} shows that the two WiSARD variants are largely unaffected by random perturbations: both retain more than 98\% of their clean accuracy under the considered random attacks. In contrast, BTHOWeN with Gaussian thermometer encoding suffers a severe
degradation, dropping from 95.20\% clean accuracy to nearly random performance. This contrast is important because BTHOWeN is the most accurate model on clean data, but also the least stable under these perturbations.

\begin{table}[t]
\centering  
\caption{Clean accuracy, post-perturbation accuracy, and retention under Random
$L_\infty$ and Random $L_2$ perturbations.}
\Description{Clean accuracy, post-perturbation accuracy, and retention under the two random attacks.}
\label{tab:random_attack}
\small
\begin{tabular}{llrrr}
\toprule
\textbf{Model} & \textbf{Probe} & \shortstack{\textbf{Clean}\\\textbf{Acc.}} & \shortstack{\textbf{Pert.}\\\textbf{Acc.}} & \textbf{Retention} \\
\midrule
DictWiSARD  & Random $L_\infty$ & 94.80\% & 94.71\% & 99.91\% \\
DictWiSARD  & Random $L_2$      & 94.80\% & 93.84\% & 98.99\% \\
Bloom WiSARD & Random $L_\infty$ & 94.75\% & 94.55\% & 99.79\% \\
Bloom WiSARD & Random $L_2$      & 94.74\% & 93.38\% & 98.56\% \\
BTHOWeN     & Random $L_\infty$ & 95.20\% & 10.12\% & 10.63\% \\
BTHOWeN     & Random $L_2$      & 95.22\% & 13.14\% & 13.80\% \\
\bottomrule
\end{tabular}
\end{table}

\subsection{Analysis of Binarization Scheme Impact}
\label{sec:binarization_impact}

\subsubsection{Linear binarization improves stability under random perturbations}

The WiSARD models use the linear thermometer of Section~\ref{sec:binarization},
whose evenly spaced thresholds make the binary code relatively stable under
small random perturbations. Returning to the 4-bit example (thresholds $0.2$,
$0.4$, $0.6$, $0.8$): for an input $x\in[0,0.1)$, no $L_\infty$ perturbation
with $\epsilon=0.1$ can alter the code, and the same holds for
$x\in(0.9,1]$; elsewhere, the number of potentially changed bits is given
exactly by Equation~\eqref{eq:threshold-crossing-bound}. With one bit per
input, only pixels within $0.1$ of the single threshold can change their code.
% Space recovery: the 4-bit example (0.2..0.8, 0.7 -> 1110) was stated verbatim
% in Section 3.1; here we only refer back to it.

\subsubsection{Gaussian binarization amplifies sensitivity to random perturbations}
The standard binarization process of the BTHOWeN model is quite different and considers the statistics of each feature from the training data to create local Gaussian thermometers. For example, if a feature has mean 0.8 and standard deviation 0.05, the generated 4-bit thermometer has values [0.75791894, 0.78733264, 0.81266736, 0.84208106]. An input $x=0.75$ is encoded as 0000, while $x+0.1 = 0.85$ is encoded as 1111.

\begin{figure}[t]
\centering
\safeincludegraphics[width=\linewidth]{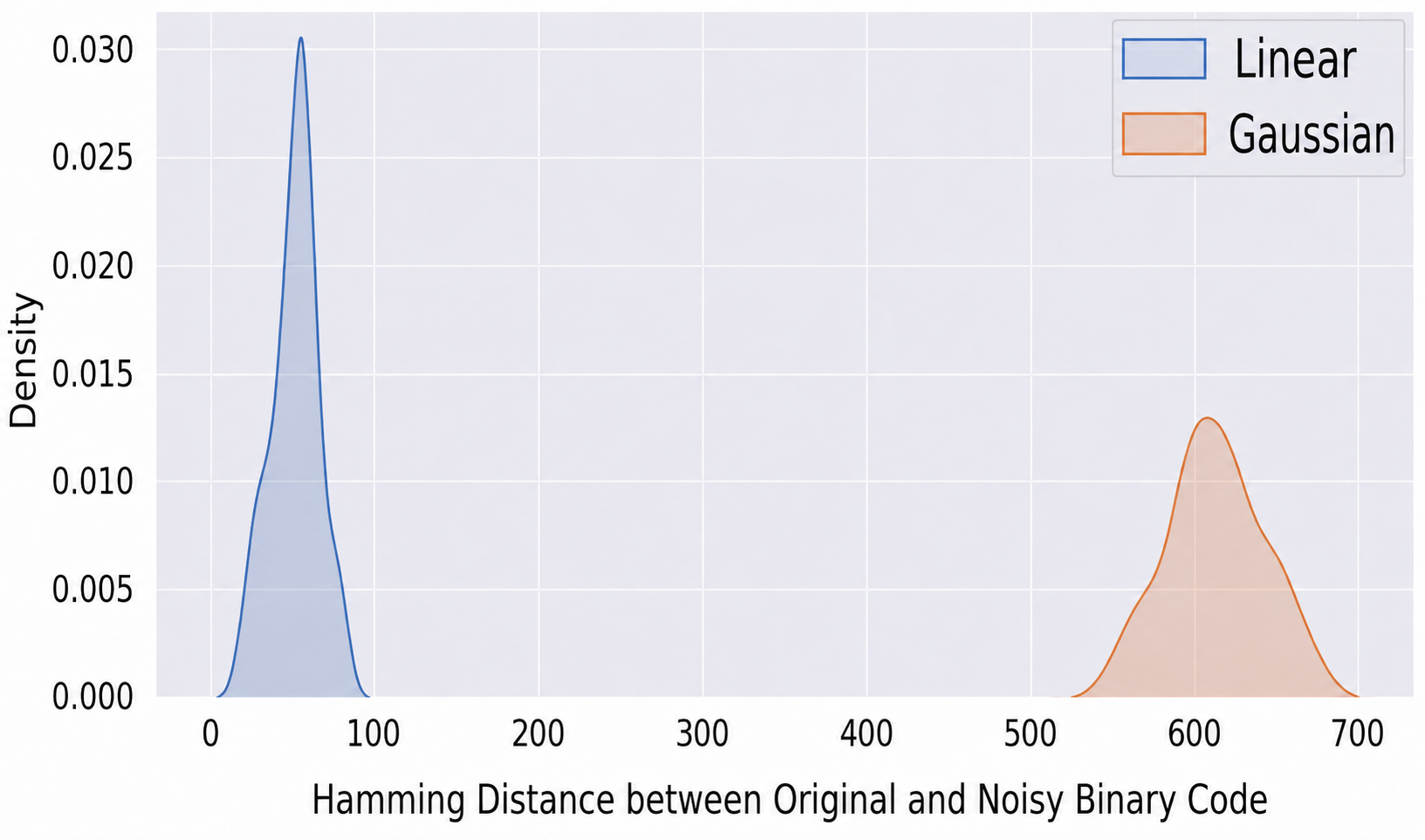}
\caption{Binarization Hamming distance between clean MNIST images and their
noisy versions under Random $L_\infty$ perturbations with $\epsilon=0.1$.
Linear binarization produces substantially smaller representation changes,
with mean Hamming distance 51.62, whereas Gaussian binarization produces a mean
Hamming distance of 613.16.}
\Description{Binarization Hamming distance between the images and their noisy versions.  The figure shows the distribution of Hamming distance ($L_{\infty}$ noise $\epsilon=0.1)$), with mean Hamming distance 51.62 and Gaussian 613.16}
\label{fig:bin_distance}
\end{figure}
\subsubsection{Comparison between linear and Gaussian-based binarization}

% Space recovery: the former opening paragraph of this subsubsection
% restated the claim of 5.4.1 and 5.4.4; the Hamming-distance measurement
% below is the actual evidence.
Figure~\ref{fig:bin_distance} quantifies the difference in representational stability between the two binarization methods on 100 test images with \texttt{bits\_per\_input} $=6$ (maximum possible Hamming distance $784\cdot 6= 4704$). The mean Hamming distance is 51.62 for linear binarization and 613.16 for Gaussian binarization, which helps explain the accuracy differences reported in Table~\ref{tab:random_attack}. %for the results shown in Table \ref{tab:random_attack}.

To confirm this, we changed the binarization method for the BTHOWeN model, while all the parameters were kept the same, including the \texttt{bits\_per\_input} = 6. The results are displayed in Table~\ref{tab:random_attack_linear_bthowen}, showing a performance retention that is much more in line with the other models. Notice that the standard Gaussian-encoded BTHOWeN model has an edge against the modified Linear-encoded BTHOWeN on the clean accuracy test (95.2\% vs 94.16\%), but whereas Gaussian-encoded BTHOWeN suffers a sharp accuracy degradation under the
same model-agnostic perturbations, Linear-encoded BTHOWeN largely preserves its performance.

\begin{table}[t]
\centering
\caption{Accuracy and retention for BTHOWeN under Random $L_\infty$ and Random $L_2$ perturbations.} \small
\Description{Accuracy and accuracy retention comparison between BTHOWeN with different binarization schemes}
\begin{tabular}{llrr}
\toprule
\textbf{BTHOWeN Model} & \textbf{Input set} & \textbf{Accuracy} & \textbf{Retention} \\
\midrule
Gaussian-encoded  & Clean set & 95.20\% & - \\
Linear-encoded   & Clean set & 94.16\% & - \\
Gaussian-encoded   & Random $L_2$   & 13.14\% & 13.80\% \\
Linear-encoded   & Random  $L_2$   & 91.88\% & 97.58\% \\
Gaussian-encoded   & Random $L_\infty$   & 10.12\% & 10.63\% \\
Linear-encoded   & Random $L_\infty$   & 94.33\% & 100.18\% \\
\bottomrule
\end{tabular}
\label{tab:random_attack_linear_bthowen}
\end{table}

\subsubsection{Comparison between norms}

Tables~\ref{tab:random_attack} and~\ref{tab:random_attack_linear_bthowen} show
that $L_\infty$ perturbations are more damaging for Gaussian binarization,
whereas $L_2$ perturbations are more effective against linear binarization.
Table~\ref{tab:attack_norms} summarizes why: the effect depends on how each
norm spreads the budget relative to how each encoding spaces its thresholds.

\begin{comment}
\subsubsection{Comparison between norms}

The relative strength of $L_\infty$ and $L_2$ perturbations depends on how each
binarization scheme places its thresholds. Tables~\ref{tab:random_attack} and
\ref{tab:random_attack_linear_bthowen} show that $L_\infty$ perturbations are
more damaging for Gaussian binarization, whereas $L_2$ perturbations are more
effective against linear binarization. This follows from the geometry of the
perturbation constraints. Under an $L_\infty$ constraint, a small change can be
applied broadly across many pixels; when thresholds are tightly packed, as in
Gaussian binarization, this can flip many bits at once. Under an $L_2$
constraint, the perturbation budget can be concentrated on fewer pixels, making
it easier to cross the more widely spaced thresholds produced by linear
thermometer encoding.
This is summarized in
Table~\ref{tab:attack_norms}. 

\end{comment}

\begin{table}[t]
\centering
\caption{Qualitative effect of $L_\infty$ and $L_2$ perturbations under
different binarization schemes.}
\Description{Qualitative comparison of how Linfinity and L2 perturbations
affect Gaussian and linear binarization.}
\label{tab:attack_norms}
\footnotesize
\setlength{\tabcolsep}{3pt}
\renewcommand{\arraystretch}{1.15}
\begin{tabular}{
    p{0.11\columnwidth}
    p{0.43\columnwidth}
    p{0.36\columnwidth}}
\toprule
\textbf{Norm} &
\textbf{Perturbation pattern} &
\textbf{More affected encoding} \\
\midrule

$L_\infty$ &
Small per-feature changes applied broadly across the input. This can flip many
bits when thresholds are close together. &
\textbf{Gaussian}, because thresholds are concentrated near high-density
regions of the data distribution. \\

\midrule

$L_2$ &
Larger changes concentrated on fewer features. This can help cross thresholds
that are farther apart. &
\textbf{Linear}, because uniformly spaced thresholds often require larger
local changes to be crossed. \\

\bottomrule
\end{tabular}
\end{table}

\subsection{Query-Based Simulated Annealing Results}
\label{subsec:sa_results}

% [CR04] ORIGINAL - Escolha da DictWiSARD no ataque adaptativo
% The previous results show that binarization strongly affects robustness under
% random perturbations. We now evaluate whether the same mechanism can be
% exploited by a query-based adversary. We focus on DictWiSARD, which was largely
% unaffected by the random attacks, and vary \texttt{bits\_per\_input} to isolate
% the effect of encoding granularity. Recall that this parameter controls how many binary
% threshold comparisons are generated from each input feature: larger values
% produce finer thermometer encodings, increasing the length of the binary
% representation and the number of thresholds that can be crossed by an
% adversarial perturbation.
% [CR04] REVISED (Ori) -- integrated into the paragraph below.
% \ori{The previous results show stability under random probes for both DictWiSARD
% and Bloom WiSARD (Table~\ref{tab:random_attack}). DictWiSARD provides an exact Boolean setting consistent with the response-margin analysis.
% Varying \texttt{bits\_per\_input} isolates encoding granularity: higher values
% create more threshold comparisons per feature and more opportunities for an
% adversarial perturbation to change the binary code.}
% [CR04] END
The random probes introduced  in Section~\ref{subsec:random_attack} and evaluated in Sections~\ref{sec:randomeval} and~\ref{sec:binarization_impact} left both WiSARD variants essentially unaffected
(Table~\ref{tab:random_attack}). We now ask whether a query-based adversary can
deliberately cross the thresholds that random noise rarely reaches. We focus
on DictWiSARD, whose exact Boolean memories match the setting of the
response-margin certificate in Section~\ref{sec:threshold_crossing_bound}, and vary
\texttt{bits\_per\_input}: higher values create more threshold comparisons per
feature, and hence more opportunities for a perturbation to change the binary
code. The attack modifies one pixel at a time, as described in
Section~\ref{subsec:sa_attack}, under the standard MNIST budget
$\epsilon=3$.

Unlike the random probes, SA places its noise where it matters, moving pixels
across thermometer thresholds; even the lowest success rate observed below,
62.0\%, is substantial.

Varying \texttt{bits\_per\_input} has only a small effect on clean accuracy,
but a large effect on adversarial vulnerability. DictWiSARD's clean accuracy
varies by less than half a percentage point, from 94.5\% to 95.0\%, as
\texttt{bits\_per\_input} goes from 1 to 6. In contrast, the Simulated
Annealing attack success rate increases from 62.0\% with one bit per input to
75.5\%, 90.0\%, and 95.0\% with 2, 4, and 6 bits per input, respectively.
These rates are computed over the first 200 test images, and images
misclassified before the attack count as successes; restricted to correctly
classified images, they are 60.2\%, 74.9\%, 89.6\%, and 94.8\%.

Thus, at this fixed budget, increasing \texttt{bits\_per\_input} provides only
marginal clean-accuracy gains but substantially increases SA success. This is
consistent with the threshold-crossing bound: finer binarization creates more
thresholds, while coarser binarization creates larger invariant regions.

\subsection{Surrogate-Transfer Results} \label{sec:surrtransfres}

%Figure~\ref{fig:surrogate-attack} in    Appendix~\ref{app:surrogate_model}  summarizes this pipeline: 

\begin{comment}
In summary, the pipeline comprises an MLP surrogate that is first trained to mimic the WNN
using soft outputs, and gradient-based adversarial examples are then generated
on the surrogate and evaluated on the original WNN.

We now evaluate the transferability of this surrogate-based attack. Instead of
relying on incidental similarity between independently trained models, this
approach explicitly trains a differentiable surrogate model to approximate the
input-output behavior of the target WNN, which is a form of model extraction via
knowledge distillation.
\end{comment}

Table~\ref{tab:transfer_results} shows the success rate of the surrogate-transfer strategy (see Section~\ref{subsec:surrogate_attack}). The ``Surrogate Fooling Rate'' is the fraction of test images correctly classified by the WNN whose adversarial version fools the MLP; it confirms that the gradient-based attacks are highly effective against the MLP itself, consistently fooling it on 89 to 95\% of these images. The key metric, ``Successful Transfer Rate,'' is the fraction of those surrogate-fooling examples that also fool the target WNN.

\begin{table}[t]
\centering \footnotesize
\caption{Adversarial transfer rate on DictWiSARD with different attacks and varying \texttt{bits\_per\_input} (bpi)}
\Description{Adversarial Transfer Rate on DictWiSARD with Different Attacks and Varying \texttt{bits\_per\_input} (bpi)}
\begin{tabular}{lcrr}
\toprule
\textbf{Attack} & \textbf{bpi} & \textbf{Successful Transfer Rate} & \textbf{Surrogate Fooling Rate} \\
\midrule
FGSM & 1 & 3.87\% & 89.32\% \\
FGSM & 2 & 7.09\% & 89.29\% \\
FGSM & 4 & 21.15\% & 91.27\% \\
FGSM & 6 & 23.61\% & 89.67\% \\
\midrule
BIM & 1 & 8.69\% & 94.32\% \\
BIM & 2 & 13.57\% & 94.59\% \\
BIM & 4 & 30.99\% & 94.46\% \\
BIM & 6 & 35.66\% & 94.86\% \\
\midrule
MI-FGSM & 1 & 7.43\% & 94.27\% \\
MI-FGSM & 2 & 15.00\% & 93.73\% \\
MI-FGSM & 4 & 25.45\% & 93.73\% \\
MI-FGSM & 6 & 39.68\% & 93.79\% \\
\midrule
PGD & 1 & 4.18\% & 93.53\% \\
PGD & 2 & 8.22\% & 93.52\% \\
PGD & 4 & 17.12\% & 94.23\% \\
PGD & 6 & 30.80\% & 93.78\% \\
\bottomrule
\end{tabular}
\label{tab:transfer_results} 
\end{table}

The transfer results follow the same qualitative trend observed with SA:
higher \texttt{bits\_per\_input} values generally increase transferability.
For example, MI-FGSM transfer rises from 7.43\% at one bit per input to 39.68\%
at six. A plausible mechanism is visible in
Equation~\eqref{eq:threshold-crossing-bound}: widely spaced thresholds leave
more surrogate perturbations inside invariant regions, whereas finer encodings
offer more opportunities to change the WNN code. This interpretation is
consistent with prior warnings that non-smooth preprocessing can create an
appearance of robustness under non-adaptive attacks~\cite{athalye2018obfuscatedgradientsfalsesense}.
% It is not proof that representation similarity causes transfer, nor that the one-bit model is robust to adaptive attacks. BIM transfers more often than FGSM in this setup, but identifying why requires a controlled surrogate architecture and fidelity ablation.

Equation~\eqref{eq:threshold-crossing-bound} also explains why transfer
success grows unevenly with \texttt{bits\_per\_input} ($b$; see
Table~\ref{tab:notation}). With linear thresholds $\tau_k=k/(b+1)$,
$k=1,\ldots,b$, and $\|\delta\|_\infty\leq 0.15$, a pixel that is $0$ before
binarization can cross a threshold only if $1/(b+1)\leq 0.15$, that is, only
from $b=6$ on, since $1/7\approx0.143$. Because 81\% of MNIST test pixels are
exactly $0$, at most 19\% of pixels can change their code when $b<6$.
Counting all pixel values, the actual fractions are 2.9\%, 6.1\%, and 10.3\%
for $b=1$, $2$, and $4$, compared with 100\% at $b=6$.

The MNIST experiments isolate the role of binarization in a controlled image
classification setting: perturbations matter when they cross encoding
thresholds and change the binary representation. In the following section, we ask whether similar
robustness issues arise in networking data, where class imbalance and the
different operational costs of false alarms and missed attacks make accuracy
alone insufficient.

\begin{table}[t]
\centering
\caption{Perturbations used in the IDS experiments.}
\label{tab:ids_attack_params}
\footnotesize
\begin{tabular}{p{1.58cm} p{1.15cm} p{1.35cm} p{3.77cm}}
\toprule
\textbf{Perturbation} & \textbf{Norm} & \textbf{Budget} & \textbf{Observation} \\
\midrule
FGSM \quad (Section~\ref{subsec:surrogate_attack})&
$L_\infty$ &
$\epsilon=0.3$ &
Single-step gradient perturbation generated on a differentiable MLP surrogate. \\
\midrule
Random $L_2$ (Section~\ref{subsec:random_attack}) &
$L_2$ &
$\epsilon=3.0$ &
Model-agnostic random perturbation. \\
\midrule
Random $L_\infty$ (Section~\ref{subsec:random_attack})  &
$L_\infty$ &
$\epsilon=0.3$ &
Model-agnostic random perturbation. \\
\bottomrule
\end{tabular}
\end{table}

\section{Intrusion-Detection Robustness}
\label{sec:ids_results}

The intrusion-detection experiments extend the MNIST analysis to a more
operational setting, where the main question is not only whether accuracy drops,
but also which types of IDS errors are induced by adversarial perturbations.

\subsection{Intrusion-Detection Datasets and Metrics }
\label{sec:ids_protocol}

% To evaluate whether the robustness patterns observed on MNIST also arise in
% networking data, we evaluate Standard WiSARD and Bloom WiSARD on Bot-IoT,
% CICIDS, and UNSW-NB15 under binary and multiclass classification. Inputs are
% preprocessed into normalized continuous feature vectors and adversarial
% perturbations are generated in this normalized feature space before WNN
% binarization. For Standard WiSARD, we vary the binarization scheme, Resolution,
% and AddressSize. For Bloom WiSARD, we additionally vary BloomSize and the
% number of hash functions. Each experimental row corresponds to an aggregated
% model configuration rather than an individual network-flow record.

To evaluate whether the robustness patterns observed on MNIST also arise in
networking data, we evaluate Standard WiSARD and Bloom WiSARD on Bot-IoT,
CICIDS, and UNSW-NB15 under binary and multiclass classification. Inputs are
preprocessed into normalized continuous feature vectors; categorical variables
are one-hot encoded and numerical variables are scaled to $[0,1]$.
We vary the binarization scheme (linear or Gaussian thermometer),
\texttt{bits\_per\_input}, and \texttt{tuple\_size}. Each table entry averages
over the configuration grid: two encodings, seven \texttt{bits\_per\_input} values
(1--32), and four \texttt{tuple\_size} values (5--20), plus, for Bloom WiSARD,
two Bloom-filter sizes and two hash counts.  % rather than an individual flow.

\textbf{IDS threat model and scope.}
We keep the evasion setting of Section~\ref{subsec:aml-background}; in the
IDS context, the attacker seeks either to move malicious traffic to the Normal
decision or to raise false alarms on benign traffic.
Table~\ref{tab:ids_attack_params} lists the three  considered perturbation methods: an FGSM transferred
from an MLP surrogate (introduced in Section~\ref{subsec:surrogate_attack}) and   two
random probes (introduced in Section~\ref{subsec:random_attack}). Perturbations are applied
to the normalized feature vector, clipped to $[0,1]$, and only then binarized.
Because one-hot groups, integer fields, and protocol-derived relations among
flow features are not enforced, perturbed vectors need not correspond to
realizable traffic; the IDS results are therefore classifier-level stress
tests rather than end-to-end evasion attacks.

\textbf{Metrics. }
For intrusion-detection experiments, accuracy alone is insufficient because the
datasets are highly imbalanced and different error types have different
operational consequences. We therefore report post-perturbation accuracy together
with Matthews Correlation Coefficient (MCC),  Attack Success
Rate (ASR), False Alarm Rate (FAR), and class-level miss rates.

\textbf{Metric definitions.}
MCC summarizes all entries of the confusion matrix and is more informative
than accuracy under severe class imbalance. ASR is the fraction of malicious
samples classified as Normal and thus measures evasion; unlike on MNIST, for the IDS experiments it is
not conditioned on samples correctly classified before the attack. FAR is the
fraction of Normal samples classified as malicious, i.e., operational false
alarms. For an intrusion class $c$, \texttt{Miss\_$c$} is the fraction of its
samples classified as Normal.

\subsection{IDS Results}

Standard WiSARD and Bloom WiSARD both degrade under perturbation. In this
section, we quantitatively assess how the extent of this degradation depends on
the dataset, task, perturbation method, and binarization scheme.

Throughout this section, a \emph{perturbation method} is how the classifier is
attacked (FGSM, Random $L_2$, or Random $L_\infty$), whereas an
\emph{intrusion class} is the network-security category being classified
(DDoS, bots, web attacks, reconnaissance, shellcode, generic traffic, and so
on). In binary tasks, \emph{Malicious} denotes the aggregate non-Normal class.

\begin{comment}
     
Figure~\ref{fig:clean_vs_adv_mcc}(a) in Appendix~\ref{app:supplementary_results} shows the relationship between clean
and post-perturbation MCC for Standard WiSARD. Many configurations with high clean MCC
move toward low or even negative post-perturbation MCC, especially in multiclass
settings. Figure~\ref{fig:clean_vs_adv_mcc}(b) shows a similar pattern for
Bloom WiSARD: several configurations with high clean MCC remain close to zero
MCC under adversarial evaluation.
\end{comment}

% Former \subsubsection{Perturbation methods used in the IDS experiments}
% (label sec:ids_perturbation_settings) removed: its content now lives in the
% threat-model paragraph of Section~\ref{sec:ids_protocol} and in the caption
% of Table~\ref{tab:ids_attack_params}.
% C\&W outputs present in preliminary experiment files are excluded because the recorded execution path could not be verified consistently across both WNN implementations.

\subsubsection{Which perturbation methods have the strongest aggregate impact} Table~\ref{tab:ids_attack_summary} summarizes the aggregate impact of the three
perturbation methods. For Standard WiSARD, FGSM causes the largest mean accuracy
drop, 35.89 percentage points, and produces the lowest mean post-perturbation MCC,
$-0.0206$. Random $L_2$ and Random $L_\infty$ also degrade performance
substantially, with mean drops of 27.02 and 24.48 percentage points,
respectively. For Bloom WiSARD, FGSM is again the most damaging perturbation
method by mean accuracy drop, with 24.58 percentage points, and also yields the
highest mean ASR, 0.5167. % Because model configurations are pooled, these means describe the evaluated grid and do not estimate uncertainty over independent training runs.

\begin{table}[t]
\centering
% \caption{Which perturbation methods have the strongest aggregate impact? Mean
% post-perturbation performance by model family and perturbation method. Lower accuracy
% drop, ASR, and FAR are better; higher MCC is better. Accuracy drop is reported
% in percentage points.}
\caption{Mean accuracy drop and post-perturbation metrics by model family and perturbation method, averaged over the linear and Gaussian configurations.}
\label{tab:ids_attack_summary}
\small
\ifrevmarks\color{black}\fi
\begin{tabular}{llrrrr}
\toprule
\textbf{Model} & \textbf{Perturbation} & \shortstack{\textbf{Acc. drop}\\\textbf{(pp)}} & \textbf{MCC} & \textbf{ASR} & \textbf{FAR} \\
\midrule
Standard & FGSM              & 35.89 & -0.0206 & 0.4786 & 0.5520 \\
Standard & Random $L_2$      & 27.02 & 0.0521  & 0.3246 & 0.6117 \\
Standard & Random $L_\infty$ & 24.48 & 0.1325  & 0.3152 & 0.5565 \\
\midrule
Bloom    & FGSM              & 24.58 & -0.0146 & 0.5167 & 0.5160 \\
Bloom    & Random $L_2$      & 15.69 & 0.0585  & 0.3681 & 0.5462 \\
Bloom    & Random $L_\infty$ & 13.54 & 0.1425  & 0.3563 & 0.5075 \\
\bottomrule
\end{tabular}
\end{table}

\subsubsection{Which operational failures do ASR and FAR reveal}
\label{sec:ids_operational_metrics}

% [CR06] ORIGINAL - Queda de acurácia versus ASR e FAR
% Beyond aggregate accuracy and MCC, IDS robustness must distinguish between two
% operational failure modes: evasion and false alarms. ASR measures evasion, i.e.,
% malicious samples classified as Normal, whereas FAR measures normal traffic
% incorrectly classified as malicious traffic. Table~\ref{tab:ids_attack_summary}
% shows that these risks do not necessarily coincide. For Standard WiSARD, Random
% $L_2$ has the highest mean FAR, 0.5617, while FGSM has the highest mean ASR,
% 0.4835, and the lowest post-perturbation MCC, $-0.0288$. For Bloom WiSARD, FGSM has
% the highest mean ASR, 0.5248, whereas Random $L_2$ has the highest mean FAR,
% 0.5050. Thus, evaluating IDS robustness requires separating perturbation methods
% that cause missed detections from those that increase false alarms.
% [CR06] REVISED (Ori) -- same content, reordered: claim, evidence, caveat.
% \ori{Accuracy drop depends on clean baselines and class prevalence, whereas ASR
% and FAR are post-perturbation malicious-to-Normal and Normal-to-malicious
% rates. Their averages pool heterogeneous configurations, so a smaller drop
% need not imply lower evasion. Table~\ref{tab:ids_attack_summary} shows higher
% ASR but lower FAR for Bloom than Standard. For Standard WiSARD, Random $L_2$
% has the highest mean FAR, 0.5617, while FGSM has the highest mean ASR,
% 0.4835, and the lowest post-perturbation MCC, $-0.0288$. For Bloom WiSARD,
% FGSM has the highest mean ASR, 0.5248, whereas Random $L_2$ has the highest
% mean FAR, 0.5050.}
Beyond aggregate accuracy and MCC, IDS robustness must separate two
operational failure modes: evasion, measured by ASR (malicious-to-Normal), and
false alarms, measured by FAR (Normal-to-malicious).
Table~\ref{tab:ids_attack_summary} shows that the two do not coincide. For
Standard WiSARD, FGSM has the highest mean ASR, 0.4786, and the lowest
post-perturbation MCC, $-0.0206$, whereas Random $L_2$ has the highest mean
FAR, 0.6117. For Bloom WiSARD, FGSM again has the highest mean ASR, 0.5167,
whereas Random $L_2$ has the highest mean FAR, 0.5462. The comparison between
families also shows why accuracy drop alone is misleading: Bloom WiSARD has
smaller mean drops than Standard WiSARD but higher ASR and lower FAR. Accuracy
drop depends on the clean baseline and on class prevalence, and these means
pool heterogeneous configurations, so a smaller drop does not imply less
evasion. Evaluating IDS robustness therefore requires reporting missed
detections and false alarms separately  (see ASR and FAR in
Table~\ref{tab:ids_attack_summary}).
% [CR06] END

% ------------------------------------------------------------------
% Surrogate-transfer methodology
% ------------------------------------------------------------------

% ------------------------------------------------------------------
% Dataset/task summary
% ------------------------------------------------------------------

% ------------------------------------------------------------------
% Norm comparison
% ------------------------------------------------------------------

% ------------------------------------------------------------------
% Miss rate by intrusion class
% ------------------------------------------------------------------

\subsubsection{Which datasets and tasks are most vulnerable}

Table~\ref{tab:ids_dataset_task_summary} aggregates results by dataset and task.
For Standard WiSARD, the most vulnerable scenario by mean accuracy drop is
Bot-IoT multiclass, with a 55.06 percentage-point drop and post-perturbation MCC
of 0.0119; UNSW-NB15 multiclass is nearly as fragile, with a 54.49
percentage-point drop and MCC of 0.0160. UNSW-NB15 binary has a negative mean
MCC, $-0.1321$, i.e., worse than chance. In contrast, Bot-IoT binary preserves
high post-perturbation accuracy on average (0.9341), but its MCC of 0.1000 shows
that this does not imply low operational risk.

For Bloom WiSARD, the lowest mean post-perturbation MCC is observed in
UNSW-NB15 binary, $-0.0525$. CICIDS multiclass is particularly informative: its
mean accuracy drop is negative ($-17.76$ percentage points), indicating that
aggregate accuracy can improve while MCC (0.0705) and ASR (0.8760) reveal poor
security behavior. CICIDS binary and multiclass exhibit high ASR for both
models, showing that evasion into the Normal class remains a central concern
even when post-perturbation accuracy appears high.

\begin{table}[t]
\centering
\setlength{\tabcolsep}{3.5pt}
% \caption{Mean adversarial performance by model family, dataset, and task.
% Accuracy drop is reported in percentage points; accuracy, MCC, and ASR are
% measured under adversarial perturbation.}
\caption{Mean post-perturbation performance by model family, dataset, and task; accuracy drop in percentage points.}
\label{tab:ids_dataset_task_summary}
\ifrevmarks\color{black}\fi
\begin{tabular}{lllrrrr}
\toprule
\textbf{Model} & \textbf{Dataset} & \textbf{Task} &
\textbf{Acc.} & \textbf{Drop} & \textbf{MCC} & \textbf{ASR} \\
\midrule
Standard & Bot-IoT   & Binary     & 0.9341 & 6.59   & 0.1000  & 0.0658 \\
Standard & Bot-IoT   & Multiclass & 0.3226 & 55.06  & 0.0119  & 0.0192 \\
Standard & CICIDS    & Binary     & 0.8418 & 9.52   & 0.2337  & 0.8517 \\
Standard & CICIDS    & Multiclass & 0.7778 & 15.65  & 0.0984  & 0.8700 \\
Standard & UNSW-NB15 & Binary     & 0.4603 & 33.46  & -0.1321 & 0.3146 \\
Standard & UNSW-NB15 & Multiclass & 0.1431 & 54.49  & 0.0160  & 0.1156 \\
\midrule
Bloom    & Bot-IoT   & Binary     & 0.8125 & 18.75  & 0.1152  & 0.1874 \\
Bloom    & Bot-IoT   & Multiclass & 0.3955 & 40.93  & 0.0144  & 0.0100 \\
Bloom    & CICIDS    & Binary     & 0.8412 & 5.69   & 0.2047  & 0.8666 \\
Bloom    & CICIDS    & Multiclass & 0.7725 & -17.76 & 0.0705  & 0.8760 \\
Bloom    & UNSW-NB15 & Binary     & 0.4846 & 27.28  & -0.0525 & 0.4325 \\
Bloom    & UNSW-NB15 & Multiclass & 0.1569 & 32.73  & 0.0206  & 0.1095 \\
\bottomrule
\end{tabular}
\end{table}

\subsubsection{How does binarization interact with perturbation method}
\label{sec:ids_binarization_hyperparams}
% Table~\ref{tab:binarization_attack_mcc} in Appendix~\ref{app:supplementary_results}
% \XXX{connects the IDS results back to the mechanism studied on MNIST by comparing the linear and Gaussian encodings under each perturbation method.}
% [CR07] ORIGINAL - Remoção da menção a resultados de C&W
% We restrict this comparison to
% the fully documented linear and Gaussian encodings and omit the unverified
% C\&W results.
% [CR07] REVISED (Ori) -- sentence dropped: with the C&W mention gone it had
% no referent ("restrict" relative to what?). The two encodings are now named
% in the lead sentence and in Section~\ref{sec:ids_protocol}.
% \ori{We restrict this comparison to the fully documented linear and Gaussian
% encodings.}
% [CR07] END 
For Standard WiSARD, linear binarization yields the higher mean
MCC under FGSM, Random $L_2$, and Random $L_\infty$, reaching 0.011, 0.086,
and 0.230, respectively, versus $-0.053$, 0.018, and 0.035 for Gaussian
binarization. For Bloom WiSARD, linear binarization likewise gives the higher mean MCC under Random $L_2$ and Random $L_\infty$  (0.094 and 0.253, versus 0.023
and 0.032), while the two encodings tie under FGSM ($-0.014$ versus $-0.015$). 

% \XXX{These are aggregate associations rather than controlled effects, since configurations also vary in \texttt{Resolution} and \texttt{AddressSize}.}

These results reinforce that the relationship between binarization and
robustness is perturbation-dependent. The same encoding strategy can have
different effects on MCC, ASR and FAR, so robustness cannot be
characterized by a single clean-accuracy or adversarial-accuracy number.  

% Additional analysis of \texttt{Resolution}, \texttt{AddressSize}, and Bloom-filter parameters is provided in Appendix~\ref{app:supplementary_results}.

\subsubsection{Which intrusion classes are most often missed as Normal}

Class-level miss rates provide a complementary view of evasion. Since
class-level miss is defined as the fraction of samples from a given intrusion
class classified as Normal, high values identify network-security categories
that are likely to bypass the detector. Table~\ref{tab:adv_miss_by_class} shows
that CICIDS is the most exposed dataset for both models: every intrusion
class is missed at rates between 0.84 and 0.91. UNSW-NB15 shows intermediate
vulnerability: binary Malicious traffic is missed at 0.31 by Standard WiSARD
and 0.43 by Bloom WiSARD, while multiclass rates range from 0.07 to 0.18 and
from 0.03 to 0.25, respectively, with generic traffic the least missed. Bot-IoT
has the smallest miss rates, at most 0.04 in the multiclass task, although
Bloom WiSARD misses 0.19 of binary Malicious traffic. Overall, these differences show that IDS robustness
must be reported not only through aggregate accuracy, but also
through class-level evasion metrics.

\begin{table}[t]
\centering
\caption{Mean adversarial miss rate by dataset, task, and intrusion class.
A miss is  an intrusion sample classified as Normal; for binary
tasks it equals the ASR in Table~\ref{tab:ids_dataset_task_summary}.}
\label{tab:adv_miss_by_class}
\small
\setlength{\tabcolsep}{3pt}
\ifrevmarks\color{black}\fi
\begin{tabular}{lllrr}
\toprule
\textbf{Dataset} & \textbf{Task} & \textbf{Class} &
\textbf{Standard} & \textbf{Bloom} \\
\midrule
CICIDS & Multiclass & Bots & 0.91 & 0.90 \\
CICIDS & Multiclass & Web attacks & 0.91 & 0.91 \\
CICIDS & Multiclass & Port scanning & 0.90 & 0.91 \\
CICIDS & Multiclass & Brute force & 0.90 & 0.90 \\
CICIDS & Multiclass & DDoS & 0.89 & 0.89 \\
CICIDS & Binary & Malicious & 0.85 & 0.87 \\
CICIDS & Multiclass & DoS & 0.84 & 0.85 \\
\midrule
UNSW-NB15 & Binary & Malicious & 0.31 & 0.43 \\
UNSW-NB15 & Multiclass & DoS & 0.18 & 0.20 \\
UNSW-NB15 & Multiclass & Analysis & 0.18 & 0.25 \\
UNSW-NB15 & Multiclass & Backdoor & 0.17 & 0.23 \\
UNSW-NB15 & Multiclass & Worms & 0.17 & 0.14 \\
UNSW-NB15 & Multiclass & Exploits & 0.15 & 0.16 \\
UNSW-NB15 & Multiclass & Fuzzers & 0.15 & 0.18 \\
UNSW-NB15 & Multiclass & Reconnaissance & 0.11 & 0.10 \\
UNSW-NB15 & Multiclass & Shellcode & 0.09 & 0.08 \\
UNSW-NB15 & Multiclass & Generic & 0.07 & 0.03 \\
\midrule
Bot-IoT & Binary & Malicious & 0.07 & 0.19 \\
Bot-IoT & Multiclass & Reconnaissance & 0.04 & 0.03 \\
Bot-IoT & Multiclass & DDoS & 0.02 & 0.01 \\
Bot-IoT & Multiclass & DoS & 0.02 & 0.01 \\
Bot-IoT & Multiclass & Theft & 0.00 & 0.00 \\
\bottomrule
\end{tabular}
\end{table}

\section{Implications} % and \XXX{Limitations}}{Implications and Limitations}}
\label{sec:engineering_implications}
% Title changed from "Engineering Implications" so that the section matches
% the roadmap in the Introduction ("discusses their implications and
% limitations") and collects, in one place, the scope caveats that Ori's
% revisions had spread over Sections 3.2, 5.1, 5.2 and 6.1.

Our results suggest that WNN deployment in security-sensitive  
settings should not tune for clean accuracy alone. Binarization is a core design
choice: increasing \texttt{bits\_per\_input}   can improve
granularity, but also creates more threshold-crossing opportunities for
adversarial perturbations. Coarser encodings may reduce sensitivity to small
perturbations, but they are not adversarial-proof, as shown by the Simulated
Annealing and surrogate-transfer experiments. For IDS applications, robustness
should therefore be assessed with MCC, ASR, FAR, and class-level miss
rates, which expose evasion into the Normal class and false alarms on benign
traffic.

These findings also matter for Bloom-filter-based networking systems. Bloom
filters are prevalent in routing, caching, filtering, and monitoring, and Bloom
WiSARD generalizes this hash-based memory principle into a learning model.
Thus, robustness should also be considered when Bloom-filter-like structures
operate in adversarial environments. Parameters such as
\texttt{tuple\_size}
affect memory use, collisions, and adversarial behavior, and
should be tuned with the binarization scheme and deployment metric.

\begin{comment}
\textbf{Limitations.}\label{sec:limitations}
Our claims are scoped as follows. The certificate of
Equation~\eqref{eq:response-margin} is proved for exact Boolean WiSARD; Bloom
collisions, counting memories, and bleaching are studied only empirically.
MNIST is a diagnostic for threshold crossings, not evidence about harder image
tasks, and only the within-DictWiSARD and BTHOWeN re-encoding experiments
isolate binarization; cross-model comparisons are not capacity matched. The
score-based attack targets only DictWiSARD and the gradient attacks are
transferred from a surrogate, so gradient masking is not ruled out. The IDS
perturbations are not constrained to realizable flows and the IDS tables pool
heterogeneous configurations, so they measure classifier sensitivity rather
than packet-level evasion.

\end{comment}

\section{Conclusion and Future Work}
\label{chap:concl}

This study identifies the precise interface through which continuous
perturbations reach a WiSARD-family classifier: encoding thresholds are
crossed, binary tuples change, and memory queries are redirected. The resulting
bound explains why coarse encodings can be locally invariant while also making
clear that discreteness is not a robustness certificate. In controlled MNIST
experiments, increasing encoding resolution substantially increases the
success of score-based and transfer attacks with little clean-accuracy gain. 
%,
% [CR08] ORIGINAL - Coerência da conclusão sobre arquitetura
% The broader cross-family results show additional interactions with hashing and
% memory design.
% [CR08] REVISED (Ori) -- kept, joined to the preceding sentence.
% \XXX{while the broader cross-family results suggest interactions with hashing and memory design without isolating their effects.}
% [CR08] END
% 
The IDS experiments reinforce a separate operational point: accuracy can hide
malicious-to-Normal errors, false alarms, and failures on minority classes.

We investigated vulnerabilities but did not yet evaluate defenses.
Several extensions follow. The response-margin
certificate can be extended to Bloom collisions, counting memories, and
bleaching. % Score-based attacks can be run against Bloom WiSARD and BTHOWeN across encoding resolutions under matched perturbation and query budgets, complemented by other adaptive black-box attacks, ablations that separate hashing, capacity, and bleaching, and harder image datasets. 
For intrusion
detection, perturbations can be restricted to attacker-controllable packet
fields, with protocol constraints enforced and derived features recomputed.
Finally, the threshold-crossing bound suggests two defenses: placing
thresholds away from the training data and threshold-aware adversarial
training. % The value of these defenses must be measured against adaptive attacks.

More broadly, in WNNs the binarizer is not an incidental preprocessing
step: it determines which perturbations the classifier can see. It should
therefore be designed, tuned, and evaluated as part of the attack surface,
and this work takes a first step toward weightless models that are robust by
design rather than by accident.

\section*{Reproducibility}
\label{sec:repro}
Code and results are available at
\url{https://github.com/lucassenosc/adversarialwnn}.

% \section*{Reproducibility}
% \label{sec:repro}
% The code and experimental reports for the intrusion-detection experiments
% (Section~6) are available in the IDS artifact.%
% \footnote{\url{https://github.com/lucassenosc/adversarialwnn}}
% PyWiSARD provides implementations of DictWiSARD and Bloom WiSARD and
% example scripts for MNIST.%
% \footnote{\url{https://github.com/theuvargas/pywisard}}
% The repository includes the Standard WiSARD and Bloom WiSARD experimental pipelines, data-preparation and aggregation utilities,  scripts, and the dependency specification required to reconstruct the experimental environment. % Large datasets are not versioned in the repository and are obtained separately following the instructions provided in its README. 

\section*{Acknowledgements}

 This work was
  supported by CAPES and FAPERJ under grants E-26/204.268/2024 and E-26/260.168/2026,
CNPq   444956/2024-7, 424622/2021-1, 315106/2023-9, and 307770/2025-7, as well as Finep PlatCiber.   Pedro Teixeira was also supported by Elytron Cybersecurity. The authors thank Gabriel Menezes  Rebello for fruitful discussions during the preparation of the camera-ready version of this paper.

 \section*{LLM Usage Statement}
LLMs were used for editorial purposes and to assist with LaTeX editing. All outputs were inspected by the authors to ensure accuracy.

\begin{comment}
    
More broadly, in WNNs the binarizer is not an incidental preprocessing
step: it determines which perturbations the classifier can see. It should
therefore be designed, tuned, and evaluated as part of the attack surface.
The study diagnoses vulnerabilities but evaluates no defense, and the
limitations above map directly onto future work: extending the certificate to
Bloom collisions, counting memories, and bleaching; running SA on Bloom WiSARD
and BTHOWeN across encoding resolutions under matched perturbation and query
budgets, together with adaptive black-box attacks reporting success and query
cost; ablating hashing, capacity, and bleaching separately and evaluating
harder image datasets; and, for IDS, restricting perturbations to
attacker-controllable packet fields with protocol constraints enforced and
derived features recomputed. As defenses, we will evaluate threshold selection
that increases training-sample distance to thresholds
(Equations~\eqref{eq:threshold-crossing-bound}--\eqref{eq:response-margin})
and threshold-aware adversarial training, measuring clean accuracy and
adaptive-attack success before making any defensive claim.
% [CR09] END
\end{comment}

\bibliographystyle{ieeetr}
\bibliography{references}

\begin{thebibliography}{10}

\bibitem{bacellar2024differentiable}
A.~T. Bacellar, Z.~Susskind, M.~Breternitz, E.~John, L.~K. John, P.~M. Lima, F.~M. Fran{\c{c}}a, {\em et~al.}, ``{Differentiable weightless neural networks},'' in {\em Proceedings of the 41st International Conference on Machine Learning, PMLR}, vol.~235, pp.~2277--2295, 2024.

\bibitem{carneiro2019exact}
H.~C. Carneiro, C.~E. Pedreira, F.~M. Fran{\c{c}}a, and P.~M. Lima, ``{The exact VC dimension of the WiSARD n-tuple classifier},'' {\em Neural computation}, vol.~31, no.~1, pp.~176--207, 2019.

\bibitem{WiSARD}
I.~Aleksander, W.~V. Thomas, and P.~Bowden, ``{WiSARD: a radical step forward in image recognition},'' {\em Sensor Review}, vol.~4, pp.~120--124, 1984.

\bibitem{8702797}
V.~C. Ferreira, A.~S. Nery, L.~A.~J. Marzulo, L.~Santiago, D.~Souza, B.~F. Goldstein, F.~M.~G. França, and V.~Alves, ``{A Feasible FPGA Weightless Neural Accelerator},'' in {\em 2019 IEEE International Symposium on Circuits and Systems (ISCAS)}, pp.~1--5, 2019.

\bibitem{BTHOWeN}
Z.~Susskind, A.~Arora, I.~D. Miranda, L.~A. Villon, R.~F. Katopodis, L.~S. De~Ara{\'u}jo, D.~L. Dutra, P.~M. Lima, F.~M. Fran{\c{c}}a, M.~Breternitz~Jr, {\em et~al.}, ``Weightless neural networks for efficient edge inference,'' in {\em Proceedings of the International Conference on Parallel Architectures and Compilation Techniques}, pp.~279--290, 2022.

\bibitem{barbosa2025updating}
R.~B. Barbosa, D.~L.~C. Dutra, P.~M.~V. Lima, D.~Carvalho, and F.~M.~G. Fran{\c{c}}a, ``{Updating KernelCanvas for weightless graph classification},'' {\em Neurocomputing}, vol.~646, p.~130458, 2025.

\bibitem{bloom_filter}
B.~H. Bloom, ``Space/time trade-offs in hash coding with allowable errors,'' {\em Commun. ACM}, vol.~13, p.~422–426, July 1970.

\bibitem{reviriego2021analyzing}
P.~Reviriego, O.~Rottenstreich, S.~Liu, and F.~Lombardi, ``{Analyzing and assessing pollution attacks on Bloom filters: Some filters are more vulnerable than others},'' in {\em 2021 17th International Conference on Network and Service Management (CNSM)}, pp.~491--499, 2021.

\bibitem{10458880}
A.~Kotzer, D.~Gandelman, and O.~Rottenstreich, ``{SoK}: Applications of sketches and rollups in blockchain networks,'' {\em IEEE Transactions on Network and Service Management}, vol.~21, no.~3, pp.~3194--3208, 2024.

\bibitem{banoun2026learning}
Y.~Banoun, D.~S. Menasch{\'e}, and O.~Rottenstreich, ``Learning filters with certainty,'' in {\em Proceedings of the 10th Asia-Pacific Workshop on Networking (APNet 2026)}, (Singapore), 2026.
\newblock Co-located with ACM SIGCOMM 2026. arXiv:2606.22786.

\bibitem{almashaqbeh2024adversary}
G.~Almashaqbeh, A.~Bishop, and H.~Tirmazi, ``{Adversary resilient learned Bloom filters},'' {\em arXiv preprint arXiv:2409.06556}, 2024.

\bibitem{lotan2025adversarially}
C.~Lotan and M.~Naor, ``{Adversarially Robust Bloom Filters: Monotonicity and Betting},'' {\em IACR Commun. Cryptol.}, vol.~2, no.~1, p.~24, 2025.

\bibitem{gerbet2015power}
T.~Gerbet, A.~Kumar, and C.~Lauradoux, ``{The power of evil choices in Bloom filters},'' in {\em 2015 45th Annual IEEE/IFIP International Conference on dependable systems and networks}, pp.~101--112, IEEE, 2015.

\bibitem{pelekis2025adversarial}
S.~Pelekis, T.~Koutroubas, A.~Blika, A.~Berdelis, E.~Karakolis, C.~Ntanos, E.~Spiliotis, and D.~Askounis, ``Adversarial machine learning: a review of methods, tools, and critical industry sectors: S. pelekis et al.,'' {\em Artificial Intelligence Review}, vol.~58, no.~8, p.~226, 2025.

\bibitem{intriguing_properties}
C.~Szegedy, W.~Zaremba, I.~Sutskever, J.~Bruna, D.~Erhan, I.~Goodfellow, and R.~Fergus, ``Intriguing properties of neural networks,'' {\em arXiv}, 2014.
\newblock \url{https://arxiv.org/abs/1312.6199}.

\bibitem{bbleaching}
D.~Carvalho, H.~Carneiro, F.~França, and P.~Lima, ``{B-bleaching: Agile Overtraining Avoidance in the WiSARD Weightless Neural Classifier},'' in {\em ESANN}, 04 2013.

\bibitem{deng2012mnist}
L.~Deng, ``{The Modified National Institute of Standards and Technology (MNIST) database of handwritten digit images for machine learning research},'' {\em IEEE signal processing magazine}, vol.~29, no.~6, pp.~141--142, 2012.

\bibitem{bloom_WiSARD}
L.~Santiago, L.~Verona, F.~Rangel, F.~Faria, D.~Menasché, W.~Caarls, M.~Breternitz, S.~Kundu, P.~Lima, and F.~França, ``{Weightless Neural Networks as Memory Segmented Bloom Filters},'' {\em Neurocomputing}, vol.~416, 02 2020.

\bibitem{bot_iot}
N.~Koroniotis, N.~Moustafa, E.~Sitnikova, and B.~Turnbull, ``{Towards the Development of Realistic Botnet Dataset in the Internet of Things for Network Forensic Analytics: Bot-IoT Dataset},'' {\em Future Generation Computer Systems}, vol.~100, pp.~779--796, 2019.

\bibitem{cicids2017}
I.~Sharafaldin, A.~Habibi~Lashkari, and A.~A. Ghorbani, ``{Toward Generating a New Intrusion Detection Dataset and Intrusion Traffic Characterization},'' in {\em Proceedings of the 4th International Conference on Information Systems Security and Privacy}, pp.~108--116, 2018.

\bibitem{unsw_nb15}
N.~Moustafa and J.~Slay, ``{UNSW-NB15: A Comprehensive Data Set for Network Intrusion Detection Systems (UNSW-NB15 Network Data Set)},'' in {\em 2015 Military Communications and Information Systems Conference}, pp.~1--6, 2015.

\bibitem{jadhao2025hybrid}
M.~Jadhao, A.~T. Bacellar, S.~Nag, I.~D. Miranda, F.~M. Fran{\c{c}}a, and L.~K. John, ``Hybrid weightless neural networks for efficient edge inference,'' in {\em 2025 35th International Conference on Field-Programmable Logic and Applications (FPL)}, 2025.

\bibitem{bacellar2026arithmetic}
A.~T. Bacellar, S.~Chemudupati, S.~Nag, A.~Seigler, P.~Lima, F.~M. Fran{\c{c}}a, and L.~K. John, ``{From Arithmetic to Logic: The Resilience of Logic and Lookup-Based Neural Networks Under Parameter Bit-Flips},'' {\em arXiv preprint arXiv:2603.22770}, 2026.

\bibitem{thermometer}
H.~Carneiro, F.~França, and P.~Lima, ``Multilingual part-of-speech tagging with weightless neural networks,'' {\em Neural Networks}, vol.~66, 03 2015.

\bibitem{buckman2018thermometer}
J.~Buckman, A.~Roy, C.~Raffel, and I.~Goodfellow, ``Thermometer encoding: One hot way to resist adversarial examples,'' in {\em International Conference on Learning Representations}, 2018.

\bibitem{madry2019deeplearningmodelsresistant}
A.~Madry, A.~Makelov, L.~Schmidt, D.~Tsipras, and A.~Vladu, ``Towards deep learning models resistant to adversarial attacks,'' {\em arXiv preprint arXiv:1706.06083}, 2017.

\bibitem{ludermir1999weightless}
T.~B. Ludermir, A.~C. P. d. L. F.~d. Carvalho, A.~P. Braga, and M.~C. De~Souto, ``Weightless neural models: a review of current and past works,'' {\em Neural Computing Surveys}, vol.~2, pp.~41--61, 1999.

\bibitem{survey_attacks_defenses}
J.~C. Costa, T.~Roxo, H.~Proença, and P.~R.~M. Inácio, ``How deep learning sees the world: A survey on adversarial attacks and defenses,'' {\em IEEE Access}, vol.~12, p.~61113–61136, 2024.

\bibitem{practical_black_box}
N.~Papernot, P.~McDaniel, I.~Goodfellow, S.~Jha, Z.~B. Celik, and A.~Swami, ``Practical black-box attacks against machine learning,'' {\em arXiv}, 2017.
\newblock \url{https://arxiv.org/abs/1602.02697}.

\bibitem{papernot2016transferabilitymachinelearningphenomena}
N.~Papernot, P.~McDaniel, and I.~Goodfellow, ``Transferability in machine learning: from phenomena to black-box attacks using adversarial samples,'' {\em arXiv}, 2016.
\newblock \url{https://arxiv.org/abs/1605.07277}.

\bibitem{counting_bloom}
L.~Fan, P.~Cao, J.~Almeida, and A.~Broder, ``Summary cache: a scalable wide-area web cache sharing protocol,'' {\em IEEE/ACM Transactions on Networking}, vol.~8, no.~3, pp.~281--293, 2000.

\bibitem{rauber2017foolbox}
J.~Rauber, W.~Brendel, and M.~Bethge, ``Foolbox: A python toolbox to benchmark the robustness of machine learning models,'' in {\em Reliable Machine Learning in the Wild Workshop, 34th International Conference on Machine Learning}, 2017.

\bibitem{goodfellow2015explainingharnessingadversarialexamples}
I.~J. Goodfellow, J.~Shlens, and C.~Szegedy, ``Explaining and harnessing adversarial examples,'' {\em arXiv preprint arXiv:1412.6572}, 2014.

\bibitem{kurakin2017adversarialexamplesphysicalworld}
A.~Kurakin, I.~Goodfellow, and S.~Bengio, ``Adversarial examples in the physical world,'' {\em arXiv preprint arXiv:1607.02533}, 2017.
\newblock https://arxiv.org/abs/1607.02533.

\bibitem{dong2018boostingadversarialattacksmomentum}
Y.~Dong, F.~Liao, T.~Pang, H.~Su, J.~Zhu, X.~Hu, and J.~Li, ``Boosting adversarial attacks with momentum,'' 2018.

\bibitem{mnist}
Y.~Lecun, L.~Bottou, Y.~Bengio, and P.~Haffner, ``Gradient-based learning applied to document recognition,'' {\em Proceedings of the IEEE}, vol.~86, no.~11, pp.~2278--2324, 1998.

\bibitem{athalye2018obfuscatedgradientsfalsesense}
A.~Athalye, N.~Carlini, and D.~Wagner, ``Obfuscated gradients give a false sense of security: Circumventing defenses to adversarial examples,'' in {\em International conference on machine learning}, pp.~274--283, PMLR, 2018.

\end{thebibliography}
 
% Keep references.bib alongside this source when compiling the complete project.
 
 \setlength{\tabcolsep}{4pt}

\end{document}